\documentclass{vldb}

\makeatletter
\def\@copyrightspace{\relax}
\makeatother

\newif\ifextended
\extendedtrue

\usepackage{subfiles}

\usepackage{graphicx}
\graphicspath{{images/}}
\DeclareGraphicsExtensions{.pdf,.jpg,.png}
\usepackage{wrapfig}
\usepackage[font=small,skip=2pt]{caption}
\usepackage{subcaption}
\usepackage{textcomp}
\usepackage{mathtools}

\usepackage{algorithm}
\usepackage[noend]{algpseudocode}%
\algnewcommand\algorithmicforeach{\textbf{for each}}
\algdef{S}[FOR]{ForEach}[1]{\algorithmicforeach\ #1\ \algorithmicdo}

\newtheorem{theorem}{Theorem}

\newtheorem{corollary}{Corollary}[theorem]
\newcommand{\pctext}[2]{\text{\parbox{#1}{\centering #2}}}
\usepackage{multirow}

\usepackage{listings} %
\lstMakeShortInline[columns=fixed]|

\usepackage{array}
\newcommand{\PreserveBackslash}[1]{\let\temp=\\#1\let\\=\temp}
\newcolumntype{C}[1]{>{\PreserveBackslash\centering}p{#1}}
\newcolumntype{R}[1]{>{\PreserveBackslash\raggedleft}p{#1}}
\newcolumntype{L}[1]{>{\PreserveBackslash\raggedright}p{#1}}
\usepackage{multirow}
\usepackage{fancyvrb} %
\usepackage{balance}  %
\PassOptionsToPackage{hyphens}{url}\usepackage{hyperref}
\usepackage{xspace}
\usepackage{booktabs} %
\usepackage[htt]{hyphenat} %

\usepackage{xpatch} %
\xapptocmd\normalsize{%
 \abovedisplayskip=5pt
 \abovedisplayshortskip=5pt
 \belowdisplayskip=5pt
 \belowdisplayshortskip=5pt
  \medskipamount=5pt %
}{}{}

\newcommand{\oursystem}{SIEVE\xspace}

\newcommand{\vDatabase}{\ensuremath{\mathcal{D}}}
\newcommand{\vRelationSet}[1]{\ensuremath{\mathcal{R}}_{#1}}
\newcommand{\vRelation}[2]{\textit{r}^{#1}_{#2}}
\newcommand{\vTupleSet}[1]{\ensuremath{\mathcal{T}}_{#1}}
\newcommand{\vTuple}[2]{\textit{t}^{#1}_{#2}}
\newcommand{\vQuery}[2]{\textit{Q}^{#1}_{#2}}

\newcommand{\vIndexSet}[1]{\ensuremath{\mathcal{I}}_{#1}}
\newcommand{\vIndex}[1]{\textit{i}_{#1}}

\newcommand{\vUserSet}[1]{\ensuremath{\mathcal{U}}_{#1}}
\newcommand{\vUser}[2]{\textit{u}^{#1}_{#2}}

\newcommand{\vGroupMethod}[1]{\textit{group(}{#1}\textit{)}}
\newcommand{\vProfileMethod}[1]{\textit{profile(}{#1}\textit{)}}

\newcommand{\vPolicySet}[1]{\ensuremath{\mathcal{P}}_{#1}}
\newcommand{\vPolicy}[2]{\textit{p}^{#1}_{#2}}

\newcommand{\vPolicyTuple}[3]{$\langle$#1, #2, #3$\rangle$}

\newcommand{\vTupleExpression}[3]{$\langle$#1, #2, #3$\rangle$}

\newcommand{\vTupleExpressionRange}[5]{$\langle$#1, #2, #3, #4, #5$\rangle$}

\newcommand{\vPolicyExpression}[1]{\textit{\ensuremath{\mathcal{E}}(}{#1}\textit{)}}

\newcommand{\vPolicyGuardedExpression}[1]{\textit{\ensuremath{\mathcal{G}}(}{#1}\textit{)}}

\newcommand{\vCostMethod}[1]{\textit{cost(}{#1}\textit{)}}

\newcommand{\vGuard}[2]{\textit{G}^{#1}_{#2}}

\newcommand{\vCandidateGuardSet}[2]{\ensuremath{\mathcal{CG}}^{#1}_{#2}}

\newcommand{\vSelectivityMethod}[1]{\rho({#1})}
\newcommand{\vReadCost}{c_r}
\newcommand{\vEvalCost}{c_e}
\newcommand{\vShortCircuit}{\alpha}
\newcommand{\vSetCardinality}[1]{\lvert {#1} \rvert}

\newcommand{\vMergeMethod}[2]{\theta({#1},{#2})}

\newcommand{\vBenefitMethod}[1]{\textit{benefit(}{#1}\textit{)}}
\newcommand{\vReadCostMethod}[1]{\textit{read\_cost(}{#1}\textit{)}}
\newcommand{\vUtilityMethod}[1]{\textit{utility(}{#1}\textit{)}}

\newcommand{\vObjectConditions}[2]{{\tt OC}^{#1}_{#2}}
\newcommand{\vObjectCondition}[2]{{\tt oc}^{#1}_{#2}}
\newcommand{\vQuerierConditions}[2]{{\tt QC}^{#1}_{#2}}
\newcommand{\vQuerierCondition}[2]{{\tt qc}^{#1}_{#2}}

\newcommand{\vPolicyAction}[2]{{\tt AC}^{#1}_{#2}}

\newcommand{\vQueryMetadata}[2]{{\tt QM}^{#1}_{#2}}

\newcommand{\vBaselineP}{$Baseline_{P}$}
\newcommand{\vBaselineI}{$Baseline_{I}$}
\newcommand{\vBaselineU}{$Baseline_{U}$}

\newcommand{\vOptimalK}{\widetilde{k}}
\newcommand{\vRateP}{r_{p}}
\newcommand{\vRateQ}{r_{q}}
\newcommand{\vRatePQ}{r_{pq}}
\newcommand{\vGuardGen}{C_G}

\newcommand{\squishlist}{
	\begin{list}{$\bullet$}
		{
			\setlength{\itemsep}{0pt}
			\setlength{\parsep}{3pt}
			\setlength{\topsep}{3pt}
			\setlength{\partopsep}{0pt}
			\setlength{\leftmargin}{1.5em}
			\setlength{\labelwidth}{1em}
			\setlength{\labelsep}{0.5em} } }

\newcommand{\squishend}{
	\end{list}  }
	
\begin{document}

\title{\oursystem: A Middleware Approach to Scalable Access Control for Database Management Systems}

\author{
\alignauthor
    Primal Pappachan\textsuperscript{*}, Roberto Yus\textsuperscript{*},~Sharad Mehrotra\textsuperscript{*},~Johann-Christoph Freytag\textsuperscript{**} \\
    \affaddr{\textsuperscript{*}UC, Irvine,\textsuperscript{**}Humboldt-Universit\"at zu Berlin}\\
    \email{\{primal, ryuspeir\}@uci.edu, sharad@ics.uci.edu, freytag@informatik.hu.berlin.de}
}

\numberofauthors{1} 

\pagestyle{plain} %

\maketitle

\begin{abstract}

Current approaches of enforcing FGAC in Database Management Systems (DBMS) do not scale in scenarios when the number of policies are in the order of thousands. This paper identifies such a use case in the context of emerging smart spaces wherein systems may be required by legislation, such as Europe's GDPR and California's CCPA, to empower users to specify who may have access to their data and for what purposes. We present \oursystem, a layered approach of implementing FGAC in existing database systems, that exploits a variety of it's features such as UDFs, index usage hints, query explain; to scale to large number of policies. Given a query, \oursystem exploits it's context to filter the policies that need to be checked. \oursystem also generates \textit{guarded expressions} that saves on evaluation cost by grouping the policies and cuts the read cost by exploiting database indices. Our experimental results, on two DBMS and two different datasets, show that \oursystem scales to large data sets and to large policy corpus thus supporting real-time access in applications including emerging smart environments.

\end{abstract}

\section{Introduction}
\label{sect:intro}

Organizations today capture and store large volumes of personal data that they use for a variety of purposes such as providing personalized services and advertisement. Continuous data capture, whether it be through sensors embedded in physical spaces to support location-based services (e.g., targeted ads and coupons), or in the form of web data (e.g., click-stream data) to learn users' web browsing model, has significant privacy implications~\cite{DBLP:journals/popets/ApthorpeHRNF19,DBLP:conf/edbt/Bertino16,DBLP:journals/jnca/SunCRSLYL17}. Regulations, such as the European General Data Protection Regulation (GDPR)~\cite{gdpr}, California Online Privacy Protection Act (CalOPPA)~\cite{caloppa} and Consumer Privacy Act (CCPA)~\cite{ccpa}, have imposed legislative requirements that control how organizations manage user data. These requirements include transparency about data collection, data minimization (both volume of data stored and duration of its retention), data retention (that requires personal data to be kept for no longer than is necessary for the purposes for which it is being processed), etc. A key requirement for organizations/services to collect and to use individual's data, is to adopt the principle of {\em choice and consent}~\cite{langheinrich2001privacy}\footnote{Currently, such organizations typically follow the principle of {\em notice} wherein they inform the user about data collection, but may not support mechanisms to seek consent.}. Until today, this requirement resulted in supporting mechanisms which allow users to opt-in/out and/or to specify data retention policies.

While such coarse level policies have sufficed for the web domain, recent work argued that as smart spaces become pervasive wherein sensors continuously monitor individuals (e.g., continuous physiological monitoring by wearable devices, location monitoring both inside and outside buildings), systems will need to empower users with finer control over who can access their data and for what purpose. Supporting such fine grained policies raises several significant challenges that are beginning to attract research attention. These challenges include policy languages suitable for representing data capture, processing, sharing and retention policies~\cite{DBLP:conf/codaspy/Panwar0WMV19} together with mechanisms for users to specify their policies within the system. This paper addresses one of such challenges: scaling enforcement of access control policies in the context of database query processing when the set of policies become a dominant factor/bottleneck in the computation due to their large number. This has been highlighted as one of the open challenges for Big Data management systems in recent surveys such as~\cite{Colombo2019}.

In the envisioned system that drives our research, data is dynamically captured from sensors and shared with people via queries based on user-specified access control policies. We describe a motivating use case of a smart campus in Section~\ref{sect:caseStudy} which shows that data involved in processing a simple analytical query might require checking against hundreds to thousands of access control policies. Enforcing that many access control policies in real-time during query execution is well beyond database systems today.
While our example and motivation is derived from the smart space and IoT setting, the need for such query processing with a large number of policies applies to many other domains. Especially, as argued before, for emerging legislatures such as GDPR that empower users to control their data. 

Today, database management systems (DBMSs) implement Fine-Grained Access Control (FGAC) by one of two mechanisms~\cite{bertino2011access}: 1)~\textit{Policy as schema} and~2)~\textit{Policy as data}. In the former case, access control policies are expressed as authorization views~\cite{rizvi2004extending}.
Then, the DBMS rewrites the query and executes it against the relevant views instead of the original data. These views allow administrators to control access to a subset of the columns and rows of a table. 
In the latter case, policies are stored in tables, just like data. The DBMS rewrites queries to include the policy predicates prior to execution~\cite{agrawal2002hippocratic,byun2005purpose,colombo2015efficient,colombo2017towards}. This mechanism allows administrators to express more fine-grained policies compared to views. Existing DBMS support both mechanisms, as they are both based on query rewriting~\cite{stonebraker1974access}, by appending policies as predicates to the |WHERE| clause of the original query. However, they are limited in the complexity of applications they can support due to the increased cost of query execution when the rewriting includes a large number of policies (e.g., appending 1K policies to a query might result in 2K extra predicates in the |WHERE| clause if each policy contains two conditions). 
Thus, scalable access control-driven query execution presents a novel challenge.

In this paper, we propose \oursystem, a general purpose middleware to support access control in DBMS that enables them to scale query processing with very large number of access control policies. It exploits a variety of features (index support, UDFs, hints) supported by modern DBMSs to scale to large number of policies.  A middleware implementation, layered on top of existing DBMS, allows us to test \oursystem independent of the specific DBMS. This is particularly useful in our case (motivated by IoT) since different systems offer different trade-offs in IoT settings as highlighted in~\cite{iotbenchmark}. The comparative simplicity of implementing the technique in middleware makes it simpler to implement yet another advantage - it allows us to explore the efficacy of different ideas instead of being constrained by the design choice of a specific system as shown in previous work such as~\cite{DBLP:journals/dase/ColomboF16}.

\oursystem incorporates two distinct strategies to reduce overhead: reducing the number of tuples that have to be checked against complex policy expressions and reducing the number of policies that need to be checked against each tuple. First, given a set of policies, it uses them to generate a set of {\em guarded expressions} that are chosen carefully to exploit best existing database indexes, thus reducing the number of tuples against which the complete and complex policy expression must be checked.

The technique for predicate simplification developed in Chaudhuri et al.~\cite{chaudhuri2003factorizing} inspired our guard selection algorithm presented in this paper.
The second strategy is inspired by pub-sub approaches such as \cite{sadoghi2011tree, gupta2004meghdoot, zhang2014opIndex, jafarpour2009mics}. Using this strategy, \oursystem reduces the overhead of dynamically checking policies during query processing by filtering policies that must be checked for a given tuple by exploiting the context present in the tuple (e.g., user/owner associated with the tuple) and the query metadata (e.g., the person posing the query i.e,. querier or their purpose). We define a policy check operator $\Delta$ for this task and present an implementation as a User Defined Function (UDF).

\oursystem combines the above two strategies in a single framework to reduce the overhead of policy checking during query execution. Thus, \oursystem adaptively chooses the best strategy possible given the specific query and policies defined for that querier based on a cost model estimation. We evaluate the performance of \oursystem using a real WiFi connectivity dataset captured in our building at UC Irvine, including connectivity patterns of over~40K unique devices/individuals. We generate a synthetic set of policies that such individuals could have defined to control access to their data by others. Our results highlight the benefit of the guarded expressions generated by \oursystem when compared to the traditional query rewrite approach for access control. Furthermore, our results show the efficiency achieved by \oursystem when processing different queries.

\noindent\textbf{Outline of the paper.} 
Section~\ref{sect:problemSetting} presents a case study of a real IoT deployment, where a large set of access control policies are expected to be defined, and reviews the relevant related works.
Section~\ref{sect:sieveApproach} formalizes the query, policy model, and the access control semantics of \oursystem.
Section~\ref{sect:guardSel} presents an algorithmic solution to generate appropriate \textit{guarded expressions}, the building block of \oursystem. 
Section~\ref{sect:implementingSieve} describes how \oursystem can be implemented in current databases.
\ifextended
Section~\ref{sect:guardMan} describes how \oursystem deals with dynamic scenarios in which the access control policy set gets updated.
\fi
Section~\ref{sect:exp} presents our experimental evaluation using a real dataset involving thousands of real individuals. 
Finally, Section~\ref{sect:conclusion} presents conclusions and future work.

\section{Problem Setting} 
\label{sect:problemSetting}

We present a case study based on a smart campus setting where there are a large number of FGAC policies specified by users on their collected data. Using this context, we review the related work and show they fall short in terms of managing and enforcing these large number of policies.

\subsection{Smart Campus Case Study} 
\label{sect:caseStudy}

We consider a motivating application wherein an academic campus supports variety of smart data services such as real-time queue size monitoring in different food courts, occupancy analysis to understand building usage (e.g., room occupancy as a function of time and events, determining how space organization impacts interactions amongst occupants, etc.), or automating class attendance and understanding correlations between attendance and grades~\cite{joon}.
While such solutions present interesting benefits, such as improving student performance~\cite{joon} and better space utilization, there are privacy challenges~\cite{pappachan2017towards} in the management of such data. This case study is based on our experience building a smart campus with variety of applications ranging from real-time services to offline analysis over the past 4 years. The deployed system, entitled TIPPERS~\cite{mehrotra2016tippers}, is in daily use in several buildings in our UC Irvine campus\footnote{More information about the system and the applications supported at \url{http://tippersweb.ics.uci.edu}}. TIPPERS at our campus captures connectivity events (i.e., logs of the connection of devices to WiFi APs) that can be used, among other purposes, to analyze the location of individuals to provide them with services. 

\begin{figure}[!htb]
    \centering
    \includegraphics[width=0.4\textwidth]{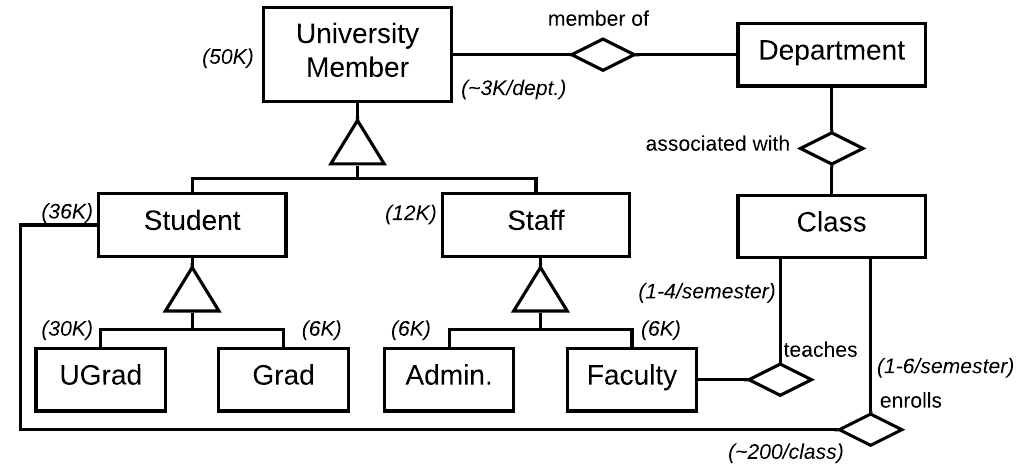}
    \caption{Entities and relationships in a Smart Campus Scenario.}
    \label{fig:entityLattice}
\end{figure}

We use the UC Irvine campus, with the various entities and relationships presented in Figure~\ref{fig:entityLattice} (along with the expected number of members in brackets), as a use case. Consider then a professor in the campus posing the following analytical query to evaluate the correlation between regular attendance in her class vs. student performance at the end of the semester:

\begin{lstlisting}
 StudentPerf(WifiDataset, Enrollment, Grades)=
 (SELECT student, grade, sum(attended)
  FROM (
   SELECT W.owner AS student, W.ts-date AS date, count(*)/count(*) AS attended
   FROM WiFiDataset AS W, Enrollment AS E
   WHERE E.class="CS101" AND E.student=W.owner AND W.ts-time between "9am" AND "10am" AND W.ts-date between "9/25/19" AND "12/12/19" AND W.wifiAP="1200"
   GROUP BY W.owner, W.ts-date) AS T, Grades AS G
  WHERE T.student=G.student
  GROUP BY T.student)
\end{lstlisting}

Let us assume that within the students in the professor's class, there exist different privacy profiles (as studied in the mobile world by Lin et al.~\cite{lin2014privacy}). Adapting the distribution of users by profile to our domain, we can assume that~20\% of the students might have a common default policy ("unconcerned" group),~18\% may want to define their own precise policies ("advance users"), and the rest will depend on the situation (for which we consider, conservatively,~2/3~to be "unconcerned" and~1/3~"advance"). Simplifying this and applying it to a class of 200 students, we have 120 unconcerned users who will adopt the default policy and 80 advanced users who will define their own set of policies. With the conservative assumption, that there are two default policies per default user and at least 4 specific policies per advanced user, we have a total of 560 policies. Typically advanced users define more policies than this conservative assumption so if we were to add two additional policy per group which will increase the number of policies to 880, or 1.2K (with three additional policies per group). 

Given the above policies for a single class, if students take 1-6 classes and faculty teach 1-4 classes per semester, a query to analyze students attendance listed above with performance over classes a professor taught over the year would be 3.3K (560 policies/class * 2 classes/quarter * 3 quarters/year) to 7.2K (considering our 1.2K policies/class estimation). We only focused on a single data type captured in this analysis (i.e., connectivity data) with two conditions per policy (e.g., time and location), and policies defined by a given user at the group-level (and not at the individual-level, which will even further increase the number of policies).

\ifextended
The case study above motivates the requirements for emerging domains, such as smart spaces and IoT, on scalable access control mechanisms for large policy sets that the DBMS must support. While our example and motivation is derived from the smart space and IoT setting, the need for such query processing with a large number of policies applies to many other domains. Especially, as argued in the introduction, for emerging legislatures such as GDPR that empower users to control their data. 
Additionally, a recent survey on future trends for access control and Big Data systems made a similar observation about the open challenge to scale policy enforcement to a large number of policies~\cite{Colombo2019}.
\fi

\subsection{Related Work}
\label{sect:relWork}

As discussed in the comprehensive survey of access control in databases in~\cite{bertino2011access}, techniques to support FGAC can be broadly classified as based on authorization views (e.g.,~\cite{rizvi2004extending} and Oracle Virtual Private Database~\cite{loney2008oracle}) or based on storing policies in the form of data (e.g., Hippocratic databases~\cite{agrawal2002hippocratic} and the follow up work~\cite{lefevre2004limiting, agrawal2005extending}). In either of these approaches, input queries are rewritten to filter out tuples for which the querier does not have access permission. This is done by adding conditions to the |WHERE| clause of the query as |$\langle$query predicate$\rangle$ AND ($P\_1$ OR ... OR $P\_n$)| (where each $P_i$ above refers to the set of predicates in each policy) or by using case-statement and outer join. Both strategies currently do not scale to scenarios with large number of policies. The view-based approach would be infeasible given the potentially large number of queriers/purposes which would result in creating and maintaining materialized views for each of them. In the policy-as-data based approach the enforcement results on computational expensive query processing. 
In a situation like the one in our use case study, it results in appending hundreds of policy conditions to the query in a disjunctive normal form which adds significant overheads.

Prior approaches, such as~\cite{byun2008purpose}, have further proposed augmenting tuples with the purpose for which they can be accessed. This reduces the overheads of policy checking at query time and could be performed at data ingestion. Such approaches have significant limitations in the context where there are large number of fine-grained polices such as in the context that motivates our work. Determining permissions for individuals and encoding them as columns or multiple rows can result in exorbitant overhead during ingestion, specially when data rates are high (e.g., hundreds of sensor observations per second). Additionally, pre-processing efforts might be wasted for those tuples that are not queried frequently or at all. Other limitations include: 1) Impossibility of pre-processing policy predicates that depend on query context or information that is not known at that time of insertion; and 2) Difficulty to deal with dynamic policies which can be updated/revoked/inserted at any time (thus requiring processing tuples already inserted when policies change). Recent work~\cite{6767117,DBLP:journals/dase/ColomboF16}, that performs some pre-processing for access control enforcement, limits pre-processing to policies explicitly defined to restrict user's access to certain queries or to certain tables. The checking/enforcement of FGAC at tuple level is deferred to query-time and enforced through query rewriting as is the case in our paper.

Several research efforts focused on access control in the context of the IoT and smart spaces. 
In~\cite{nehme2013fence}, the authors propose a policy based access control approach for sensor data. In their context, the system receives data and their associated policies based on queries submitted to the system. The approach does not handle analytical queries with policies on the arriving data. The implementation of their approach requires significant modification to existing DBMS to make different operators \textit{security-aware} for a large number of policies. 
In~\cite{colombo2018access}, the authors proposed a new architecture based on MQTT for IoT ecosystems.
The focus of these works, however, is not on managing large number of policies at run time and hence, they would experience the same issues highlighted for traditional query rewrite strategies.

\section{\oursystem approach to FGAC}
\label{sect:sieveApproach}

We describe the three fundamental entities in policy drive data processing: data, query, and policies. For policies we delve deeper and explain what each attribute in the policy is for. Then using these three components we describe the access control semantics used in this paper. We finish the section with a sketch of the approach followed by \oursystem to speed up policy enforcement. We have summarize frequently used notations are summarized in Table~\ref{table:notation_table} for perusal.

\subsection{Modeling Policy Driven Data Processing}
\label{sect:modeling}

\begin{table}[t]
\scriptsize
\centering
\begin{tabular}{m{3cm}m{5cm}}
  \textbf{Notation} & \textbf{Definition} \\ 
 \hline \\[-1em]
  $\vDatabase$  & Database \\ 
 \hline \\[-1em]
 $\vIndex{i} \in \vIndexSet{}$ & Index and set of indexes in $\vDatabase$ \\ 
 \hline \\[-1em]
 $\vRelation{}{i} \in \vRelationSet{}$ & Relation and set of relations in $\vDatabase$ \\ 
 \hline \\[-1em]
 $\vUser{}{k} \in \vUserSet{}$  & User and set of users in $\vDatabase$ \\ 
 \hline \\[-1em]
 $\vTuple{}{j} \in \vTupleSet{}$; $\vTupleSet{\vRelation{}{i}}$; $\vTupleSet{\vUser{}{k}}$;  $\vTupleSet{\vQuery{}{i}}$; $\vTupleSet{\vPolicy{}{l}}$ & Tuple and set of tuples: in $\vDatabase$; in $\vRelation{}{i}$; owned by $\vUser{}{k}$; required to compute $\vQuery{}{i}$; controlled by $\vPolicy{}{l}$ \\
 \hline \\[-1em]
 $\vGroupMethod{\vUser{}{k}}$  & Groups $\vUser{}{k}$ is part of \\ 
 \hline \\[-1em]
 $\vQuery{}{i}$; $\vQueryMetadata{i}{}$ & Query; Metadata of $\vQuery{}{i}$ \\ 
 \hline \\[-1em]
 $\vPolicy{}{l} \in \vPolicySet{}$; $\vPolicySet{\vQueryMetadata{i}{}}$ & Access control policy and set of policies in $\vDatabase$; set of policies related to a query given its metadata \\
 \hline \\[-1em]
 $\vObjectCondition{l}{i} \in \vObjectConditions{l}{}$;$\vQuerierCondition{l}{i} \in \vQuerierConditions{l}{}$;$\vPolicyAction{l}{}$ & Object conditions; querier conditions; action of $\vPolicy{}{l}$ \\ 
 \hline \\[-1em]
  $\vPolicyExpression{\vPolicySet{}} = \vObjectConditions{1}{} \lor \cdots \lor \vObjectConditions{\vSetCardinality{\vPolicySet{}}}{}$ & {\em Policy expression} of $\vPolicySet{}$ \\ 
 \hline \\[-1em]
  $\vPolicyGuardedExpression{\vPolicySet{}}=\vGuard{}{1} \lor \cdots \lor \vGuard{}{n}$ & {\em Guarded policy expression} of $\vPolicySet{}$ (DNF of guarded expressions) \\ 
 \hline \\[-1em]
  $\vGuard{}{i} = \vObjectCondition{i}{g} \land \vPolicySet{\vGuard{}{i}}$ & {\em Guarded expression} which consists of conjunctive expression of {\em guard} ($\vObjectCondition{i}{g}$) and a set of policies for which ($\vObjectCondition{i}{g}$) is a common factor. We refer to these policies as a {\em policy partition} ($\vPolicySet{\vGuard{}{i}}$) \\
 \hline \\[-1em]
  $\vCandidateGuardSet{}{}$ & {\em Candidate guards} for $\vPolicyExpression{\vPolicySet{}}$ \\
 \hline \\[-1em]
  $eval(exp, \vTuple{}{t})$ & function which evaluates a tuple $\vTuple{}{t}$ against a expression $exp$ \\
 \hline \\[-1em]
  $\Delta(\vPolicySet{\vGuard{}{i}}, \vQueryMetadata{i}{}, \vTuple{}{t})$ & policy operator \\
 \hline \\[-1em]
 $\vSelectivityMethod{pred}$  & estimated cardinality of a predicate \\
 \hline \\[-1em]
 $\vEvalCost$; $\vReadCost$  & cost of evaluating a tuple against the set of object conditions of a policy;  cost of reading a tuple from the disk  \\
 \hline \\[-1em]
 $\vShortCircuit$  & average number of policies that a tuple is checked against before it satisfies one  \\
 \end{tabular}
\caption{Frequently used notations.}
\label{table:notation_table}
\vspace{-0.5cm}
\end{table}

\vspace{0.1cm}
\noindent
\textbf{Data Model.} Let us consider a database $\vDatabase$ consisting of a set of relations $\vRelationSet{}$, a set of data tuples $\vTupleSet{}$, a set of indexes $\vIndexSet{}$, and set of users $\vUserSet{}$. $\vTupleSet{\vRelation{}{i}}$ represents the set of tuples in the relation $\vRelation{}{i} \in \vRelationSet{}$. Users are organized in collections or {\em groups}, which are hierarchical (i.e., a group can be subsumed by another). For example, the group of undergraduate students is subsumed by the group of students. Each user can belong to multiple groups and we define the method $\vGroupMethod{\vUser{}{k}}$ which returns the set of groups $\vUser{}{k}$ is member of. Each data tuple $\vTuple{}{j} \in \vTupleSet{}$ belongs to a $\vUser{}{k} \in \vUserSet{}$ or a group whose access control policies restrict/grant access over that tuple to other users. We assume that for each data tuple $\vTuple{}{j} \in \vTupleSet{}$ there exists an owner $\vUser{}{k} \in \vUserSet{}$ who owns it, whose access control policies restrict/grant access over that tuple to other users (the ownership can be also shared by users within a group). This ownership is explicitly stated in the tuple by using the attribute $\vRelation{}{i}.owner$ that exists for all $\vRelation{}{i} \in \vRelationSet{}$ and that we assume is indexed (i.e., $\forall \; \vRelation{}{i} \in \vRelationSet{} \; \exists \; \vIndex{j} \in \vIndexSet{} \mid \vIndex{j}$ is an index over the attribute $\vRelation{}{i}.owner$). $\vTupleSet{\vUser{}{k}}$ represents the set of tuples owned by user $\vUser{}{k}$.

\vspace{0.1cm}
\noindent
\textbf{Query Model.} 
The SELECT-FROM-WHERE query posed by a user $\vUser{}{k}$ is denoted by $\vQuery{}{i}$ and tuples in the relations in the FROM statement(s)  of query are denoted by $\vTupleSet{\vQuery{}{i}} = \bigcup\limits_{i=1}^{n} \vTupleSet{\vRelation{}{i}}$. In our model, we consider that queries have associated metadata $\vQueryMetadata{i}{}$ which consists of information about the querier and the context of the query. This way, we assume that for any given query $\vQuery{}{i}$, $\vQueryMetadata{i}{}$ contains  the identity of the querier (i.e., $\vQueryMetadata{i}{querier}$) as well as the purpose of the query (i.e., $\vQueryMetadata{i}{purpose}$). In the example query in Section~\ref{sect:caseStudy}, $\vQueryMetadata{i}{querier}$=``Prof.Smith" and $\vQueryMetadata{i}{purpose}$=``Analytics".

\vspace{0.1cm}
\noindent
\textbf{Access Control Policy Model.} A user specifies an access control policy (in the rest of the paper we will refer to it simply as policy) to allow or to restrict access to certain data she owns, to certain users/groups under certain conditions. Let $\vPolicySet{}$ be the set of policies defined over $\vDatabase$ such that $\vPolicy{}{l} \in \vPolicySet{}$ is defined by a user $\vUser{}{k}$ to control access to a set of data tuples in $\vRelation{}{i}$. Let that set of tuples be $\vTupleSet{\vPolicy{}{l}}$ such that $\vTupleSet{\vPolicy{}{l}} \subseteq \vTupleSet{\vUser{}{k}} \cap \; \vTupleSet{\vRelation{}{i}}$. We model such policy as $\vPolicy{}{l}=$\vPolicyTuple{$\vObjectConditions{l}{}$}{$\vQuerierConditions{l}{}$}{$\vPolicyAction{l}{}$}, where each element represents:

\noindent
$\bullet$
Object Conditions ($\vObjectConditions{l}{}$) are defined using a conjunctive boolean expression $\vObjectCondition{l}{1} \land \vObjectCondition{l}{2} \land ... \land \vObjectCondition{l}{n}$ which determines the access controlled data tuple(s).
Each {\em object condition} ($\vObjectCondition{l}{c}$) is a boolean expression \vTupleExpression{$attr$}{$op$}{$val$} where $attr$ is an attribute (or column) of $\vRelation{}{i}$, $op$ is a comparison operator (i.e., $=$, $!=$, $<$, $>$, $\geq$, $\leq$, |IN|, |NOT IN|, |ANY|, |ALL|), and $val$ can be either: (1) A constant or a range of constants or (2) A derived value(s) defined in terms of the expensive operator (e.g., a user defined function to perform face recognition) or query on $\vDatabase$ that will obtain such values when evaluated. To represent boolean expressions involving a range defined by two comparison operators (e.g., $4\leq a <20$) we use the notation \vTupleExpressionRange{$attr$}{$op1$}{$val1$}{$op2$}{$val2$} (e.g., \vTupleExpressionRange{$a$}{$\geq$}{$4$}{$<$}{$20$}). 
We assume that there exists exactly one $\vObjectCondition{l}{c} \in \vObjectConditions{l}{}$ such that $\vObjectCondition{l}{c} =$ \vTupleExpression{$\vRelation{}{i}.owner$}{=}{$\vUser{}{k}$} or $\vObjectCondition{l}{c} =$ \vTupleExpression{$\vRelation{}{i}.owner$}{=}{$\vGroupMethod{\vUser{}{k}}$}. We will refer to this object condition as $\vObjectCondition{l}{owner}$ in the rest of the paper.

\noindent
$\bullet$
Querier Conditions ($\vQuerierConditions{l}{}$) identifies the metadata attributes of the query to which the access control policy applies. $\vQuerierConditions{l}{}$ is a conjunctive boolean expression $\vQuerierCondition{l}{1} \land \vQuerierCondition{l}{2} \land \cdots \land \vQuerierCondition{l}{m}$. 
Our model follows the well studied Purpose Based Access Control (Pur-BAC) model~\cite{byun2005purpose} to define the querier conditions.
Thus, we assume that each policy contains has at least two querier conditions such as $\vQuerierCondition{l}{querier}$ =  
\vTupleExpression{$\vQueryMetadata{i}{querier}$}{=}{$\vUser{}{k}$}
or $\vQuerierCondition{l}{querier} =$ \vTupleExpression{$\vQueryMetadata{i}{querier}$}{=}{$\vGroupMethod{\vUser{}{k}}$} (that defines either a user or group), and a $\vQuerierCondition{l}{purpose}$ = \vTupleExpression{$\vQueryMetadata{i}{purpose}$}{=}{\textit{purpose}} which models the intent/purpose of the querier (e.g., safety, commercial, social, convenience, specific applications on the scenario, or any~\cite{lee2017privacy}). 
Other pieces of querier context (such as the IP of the machine from where the querier posed the query, or the time of the day) can easily be added as querier conditions although in the rest of the paper we focus on the above mentioned querier conditions.

\noindent
$\bullet$
Policy Action ($\vPolicyAction{l}{}$) defines the enforcement operation, or {\em action}, which must be applied on any tuple $\vTuple{}{j} \in \vTupleSet{\vPolicy{}{l}}$.
We consider the default action, in the absence of an explicit-policy allowing access to data, to be \textit{deny}. Such a model is standard in systems that collect/manage user data. 
Hence, explicit access control actions associated with policies in our context are limited to \textit{allow}.  
If a user expresses a policy with a deny action (e.g., to limit the scope/coverage of an allow policy), we can factor in such a deny policy into the explicitly listed allow policies. For instance, given an explicit allow policy ``allow John access to my location'' and an overlapping deny policy ``deny everyone access to my location when in my office'', we can factor in the deny policy by replacing the original allow policy by ``allow John access to my location when I am in locations other than my office''. We therefore restrict our discussions to allow policies.

Based on this policy model, we show two sample policies in the context of the motivating scenario explained before. First, we describe a policy with object conditions containing a constant value. This policy is defined by John to regulate access to his connectivity data to Prof. Smith only if he is located in the classroom and for the purpose of class attendance as follows: |$\langle$[W.owner = John $\land$ W.ts-time $\geq$ 09:00 $\land$ W.ts-time $\leq$ 10:00 $\land$ W.wifiAP = 1200], [Prof. Smith $\land$ Attendance Control], allow$\rangle$|. Second, we describe the same policy with an object condition derived from a query to express that John wants to allow access to his location data only when he is with Prof. Smith. The object condition is updated as: |[W.owner = John $\land$ W.wifiAP = (SELECT W2.wifiAP FROM WifiDataset AS W2 WHERE W2.ts-time = W.ts-time AND W2.owner = "Prof.Smith")]|

\vspace{0.1cm}
\noindent

\textbf{Access Control Semantics.} 
We define access control as the task of deriving $\vTupleSet{\vQuery{}{i}}' \subseteq \vTupleSet{\vQuery{}{i}}$ which is the projection of $\vDatabase$ on which $\vQuery{}{i}$ can be executed with respect to access control policies defined for it's querier. Thus $\forall$ $\vTuple{}{t} \in \vTupleSet{\vQuery{}{i}}$, $\vTuple{}{t} \in \vTupleSet{\vQuery{}{i}}'$ $\Leftrightarrow$ $eval(\vPolicyExpression{\vPolicySet{}}, \vTuple{}{t})=True$. The function $eval(\vPolicyExpression{\vPolicySet{}}, \vTuple{}{t})$ evaluates a tuple $\vTuple{}{t}$ against the policy expression $\vPolicyExpression{\vPolicySet{}}$ that applies to $\vQuery{}{i}$ as follows:
{\small
\begin{alignat*}{2}
  eval(\vPolicyExpression{\vPolicySet{}}, \vTuple{}{t})   &= \begin{cases}
        True {\small \pctext{2.2in}{if $\exists$ \; $\vPolicy{}{l} \in \vPolicySet{} \mid$ eval($\vObjectConditions{l}{}, \vTuple{}{t}$) = True}}%
        \\
        False {\small \pctext{2.2in}{otherwise}}%
        \\
        \end{cases}
\end{alignat*}
}%

\noindent where $eval(\vObjectConditions{l}{}, \vTuple{}{t})$ evaluates the tuple against the object conditions of $\vPolicy{}{l}$ as follows:
{\small
\begin{alignat*}{2}
  eval(\vObjectConditions{l}{}, \vTuple{}{t})   &= \begin{cases}
        True {\small \pctext{2.2in}{if $\forall \; \vObjectCondition{l}{c} \in \vObjectConditions{l}{} \mid \vTuple{}{t}$.attr = $\vObjectCondition{l}{c}$.attr $\implies$ eval($\vObjectCondition{l}{c}$.op,$\vObjectCondition{l}{c}$.val,$\vTuple{}{t}$.val) = True}}%
        \\
        False {\small \pctext{2.2in}{otherwise}}%
        \\
        \end{cases}
\end{alignat*}
}%

\noindent where $eval(\vObjectCondition{l}{c}.op,\vObjectCondition{l}{c}.val,\vTuple{}{t}.val)$ compares the object condition value ($\vObjectCondition{l}{c}.val$) to the corresponding tuple value ($\vTuple{}{t}.val$) that matches the attribute of the object condition, using the object condition operator. If the latter is a derived value, the expensive operator/query is evaluated to obtain the value. 

Given the above semantics, the order of evaluating policies and query predicates is important for correctness of results. Depending upon the query operations, evaluating policies after them is not guaranteed to produce correct results. This is trivially true in the case for aggregation or projection operations that remove certain attributes from a tuple. In queries with non-monotonic operations such as set difference, performing query operations before policy evaluation will result in inconsistent answers.

Let $\vPolicySet{}$ be the set of policies defined on $\vRelation{}{k}$ that control access to $\vQuery{}{i}$. $\vPolicyExpression{\vPolicySet{}}$ is the Disjunctive Normal Form (DNF) expression of $\vPolicySet{}$ such that $\vPolicyExpression{\vPolicySet{}} = \vObjectConditions{1}{} \lor \cdots \lor \vObjectConditions{\vSetCardinality{\vPolicySet{}}}{}$ where $\vObjectConditions{l}{}$ is conjunctive expression of object conditions from $\vPolicy{}{l} \in \vPolicySet{}$. After appending $\vPolicyExpression{\vPolicySet{}}$ to $\vQuery{}{i}$ we obtain: |SELECT * FROM $\vRelation{}{j}$ MINUS SELECT * FROM $\vRelation{}{k}$ WHERE $\vPolicyExpression{\vPolicySet{}}$|. Consider a tuple $\vTuple{}{k} \in \vTupleSet{\vRelation{}{k}}$ which has policy $\vPolicy{}{l} \in \vPolicySet{}$ that denies access $\vQuery{}{i}$ to $\vTuple{}{k}$. If there exists a tuple $\vTuple{}{j} \in \vTupleSet{\vRelation{}{j}}$ such that $\vTuple{}{j} = \vTuple{}{k}$, then performing set difference operations before checking policies on $\vRelation{}{k}$ will result in a tuple set that doesn't include $\vTuple{}{j}$. On the other hand, if policies for $\vRelation{}{k}$ are checked first, then $\vTuple{}{k} \not\in \vTupleSet{\vQuery{}{i}}$ and therefore $\vTuple{}{j}$ will be in the query result. 

This access control semantics satisfies the sound and secure properties of the correctness criterion defined by~\cite{wang2007correctness}. If no policies are defined on $\vTuple{}{t}$ then the tuple is not included in $\vTupleSet{\vQuery{}{i}}'$ as our access control semantics is opt-out by default.

\subsection{Overview of \oursystem Approach}
\label{sect:ourApproach}

For a given query $\vQuery{}{i}$, the two main factors that affect the time taken to evaluate the set of policies for the set of tuples  $\vTupleSet{\vQuery{}{i}}$ required to compute $\vQuery{}{i}$
(i.e.,  $eval(\vPolicyExpression{\vPolicySet{}}, \vTuple{}{t}) \; \forall \; \vTuple{}{t} \in \vTupleSet{\vQuery{}{i}}$) are the large number of complex policies and the number of tuples in  $\vTupleSet{\vQuery{}{i}}$.  The overhead of policy evaluation can thus be reduced by first eliminating tuples using low cost filters before checking the relevant ones against complex policies and second by minimizing the length of policy expression a tuple $\vTuple{}{t}$ needs to be checked against before deciding whether it can be included in the result of $\vQuery{}{i}$ or not. These two fundamental building blocks form the basis for \oursystem.

\squishlist

    \item \textbf{Reducing Number of Policies.}
    Not all policies in $\vPolicySet{}$ are relevant to a specific query $\vQuery{}{i}$. We can first easily filter out those policies that are defined for different queriers/purposes given the query metadata $\vQueryMetadata{i}{}$. For instance, when Prof. Smith poses a query for grading, only the policies defined for him and the faculty group for grading purpose are relevant out of all policies defined on campus.
    We denote the subset of policies which are relevant given the query metadata $\vQueryMetadata{i}{}$ by $\vPolicySet{\vQueryMetadata{i}{}} \subseteq \vPolicySet{}$ where $\vPolicy{}{l} \in \vPolicySet{\vQueryMetadata{i}{}}$ iff  $\vQueryMetadata{i}{purpose} = \vQuerierCondition{l}{purpose} \land (\vQueryMetadata{i}{querier} = \vQuerierCondition{l}{querier} \lor \vQuerierCondition{l}{querier} \in group(\vQueryMetadata{i}{querier}))$. 
    In addition, for a given tuple $\vTuple{}{t} \in \vTupleSet{\vQuery{}{i}}$ we can further filter policies in $\vPolicySet{\vQueryMetadata{i}{}}$ that we must check based on the values of attributes in $\vTuple{}{t}$.
For instance, we can further restrict the set of policies relevant for Prof. Smith's query by considering information of each tuple involved in the query such as its owner (i.e., {$\vTuple{}{t}.owner$}). This way, if the tuple belongs to John, only policies defined by John have to be checked from the previous set.

    \item \textbf{Reducing Number of Tuples.}
    Even if the number of policies to check are minimized, the resulting expression $\vPolicyExpression{\vPolicySet{}}$ might still be computationally complex. 
    We might improve performance by filtering out tuples based on low cost filters derived from $\vPolicyExpression{\vPolicySet{}}$. 
    Such  processing can be even faster if such simplified expressions could leverage  existing indexes $\vIndexSet{}$ over attributes in the database.
    We therefore rewrite the policy expression  $\vPolicyExpression{\vPolicySet{}}= \vObjectConditions{1}{} \lor \cdots \lor \vObjectConditions{\vSetCardinality{\vPolicySet{}}}{}$ as a {\em guarded policy expression} $\vPolicyGuardedExpression{\vPolicySet{}}$ which is a disjunction of {\em guarded expressions} $\vPolicyGuardedExpression{\vPolicySet{}} = \vGuard{}{1} \lor \cdots \lor \vGuard{}{n}$. Each $\vGuard{}{i}$ consists of a {\em guard} $\vObjectCondition{i}{g}$ and a {\em policy partition} $\vPolicySet{\vGuard{}{i}}$ where $\vPolicySet{\vGuard{}{i}} \subseteq \vPolicySet{}$. Note that $\vPolicySet{\vGuard{}{i}}$ partitions the set of policies, i.e., $\vPolicySet{\vGuard{}{i}} \cap \vPolicySet{\vGuard{}{j}} = \emptyset \; \forall \; \vGuard{}{i}, \vGuard{}{j} \in \vPolicyGuardedExpression{\vPolicySet{}}$. Also, all policies in $\vPolicySet{}$ are covered by one of the guarded expressions, i.e.,  $\forall \; \vPolicy{}{i} \in \vPolicySet{} \; (\exists \; \vGuard{}{i} \in \vGuard{}{} \mbox{\ such that\ } \vPolicy{}{i} \in \vPolicySet{\vGuard{}{i}})$.  We will represent the guarded expression $\vGuard{}{i} = \vObjectCondition{i}{g} \land \vPolicySet{\vGuard{}{i}}$ where $\vPolicySet{\vGuard{}{i}}$ is the set of policies but for simplicity of expression we will use it as an expression where there is a disjunction between policies.

    The {\em guard} term $\vObjectCondition{i}{g}$ is an object condition 
    that can support efficient filtering by exploiting an index. In particular, 
    it satisfies the following properties:

    \squishlist
        \item $\vObjectCondition{i}{g}$ is a simple predicate over an attribute
        (e.g., $ts-time > 9am$) consisting of an attribute name, a comparison operator, and a constant value. Also, the attribute in $\vObjectCondition{i}{g}$ has an index on it (i.e., $\vObjectCondition{i}{g}.attr \in \vIndexSet{}$).
        \item The guard  $\vObjectCondition{i}{g}$  can serve as a filter for all the policies in  the partition $\vPolicySet{\vGuard{}{i}}$ (i.e.,
        $\forall \, \vPolicy{}{l} \in \vPolicySet{\vGuard{}{i}} \; \exists \; \vObjectCondition{l}{j} \in \vObjectConditions{l}{} \mid \vObjectCondition{l}{j} \implies \vObjectCondition{i}{g}$).
    \squishend
    
\squishend

As an example, consider the policy expression of all the policies defined by students to grant the professor access to their data in different situations. Let us consider that many of such policies grant access when the student is connected to the WiFi AP of the classroom. For instance, in addition to John's policy defined before, let us consider that Mary defines the policy |$\langle$[W.owner = Mary $\land$ $\land$ W.wifiAP = 1200], [Prof. Smith $\land$ Attendance Control], allow$\rangle$|. This way, such predicate (i.e., wifiAP=1200) could be used as a guard that will group those policies, along with others that share that predicate, to create the following expression:
|wifiAP=1200 AND ((owner=John AND ts-time between 9am AND 10am OR (owner=Mary) OR ...)|

\oursystem adaptively selects a query execution strategy when a query is posed leveraging the above ideas. First, given $\vQuery{}{i}$, \oursystem filters out policies based on $\vQueryMetadata{i}{}$. Then, using the resulting set of policies it replaces any relation $\vRelation{}{j} \in \vQuery{}{i}$ by a projection that satisfies policies in $\vPolicySet{\vQueryMetadata{i}{}}$ that are defined over $\vRelation{}{j}$. It does so by using the guarded expression  $\vPolicyGuardedExpression{\vPolicySet{\vRelation{}{j}}}$ constructed as a query |SELECT * FROM $\vRelation{}{j}$ WHERE $\vPolicyGuardedExpression{\vPolicySet{\vRelation{}{j}}}$|

By using $\vPolicyGuardedExpression{\vPolicySet{\vRelation{}{j}}}$ and its guards $\vObjectCondition{i}{g}$, we can efficiently filter out a high number of tuples and only evaluate the relevant tuples against the more complex policy partitions $\vPolicySet{\vGuard{}{i}}$. The generation of $\vPolicyGuardedExpression{\vPolicySet{\vRelation{}{j}}}$ can be performed either offline before queries arrive or even online as the algorithm (see Section~\ref{sect:guardSel}) is efficient for large numbers of policies (as we will show in our experiments).

A tuple that satisfies the guard $\vObjectCondition{i}{g}$ is then checked against $\vPolicyExpression{\vPolicySet{\vGuard{}{i}}}=\vObjectConditions{1}{} \lor \cdots \lor \vObjectConditions{\vSetCardinality{\vPolicySet{\vGuard{}{i}}}}{}$. As it is a DNF expression, in the worst case (a tuple that does not satisfy any policy) will have to be evaluated against each $\vObjectConditions{j}{} \in \vPolicyExpression{\vPolicySet{\vGuard{}{i}}}$. We introduce a policy operator ($\Delta(\vPolicySet{\vGuard{}{i}}, \vQueryMetadata{i}{}, \vTuple{}{t})$) which takes $\vPolicySet{\vGuard{}{i}}$ and each tuple $\vTuple{}{t}$ that satisfied $\vGuard{}{i}$ and retrieves $\hat{\vPolicySet{\vGuard{}{i}}}$, the relevant policies to be evaluated based on $\vQueryMetadata{i}{}$ and $\vTuple{}{t}$. Then, it performs $eval(\hat{\vPolicySet{\vGuard{}{i}}}, \vTuple{}{t})$. Hence, $\vGuard{}{i}=\vObjectCondition{i}{g} \land \vPolicySet{\vGuard{}{i}}$ will become $\vGuard{}{i}=\vObjectCondition{i}{g} \land \Delta(\vPolicySet{\vGuard{}{i}}, \vQueryMetadata{i}{}, \vTuple{}{t})$.
\oursystem decides in which situations and for which specific $\vGuard{}{i} \in \vPolicyGuardedExpression{\vPolicySet{\vRelation{}{j}}}$ using the $\Delta$ operator can minimize the execution cost by estimating its cost versus the cost of $eval(\vPolicyExpression{\vPolicySet{\vGuard{}{i}}}, \vTuple{}{t})$. The details of implementation and usage of $\Delta$ are explained in Section~\ref{sect:implementingSieve}.

Hence, the main challenges are: 1) Selecting appropriate guards and generating the guarded expression; 2) Dynamically selecting a strategy and constructing a query that can be executed in an existing DBMS, using the selected strategy. We explain our algorithm to generate guarded expressions for a set of policies in Section~\ref{sect:guardSel}. 
This generation might take place offline if the policy dataset is deemed to undergo small number of changes over time. Otherwise, the generation can be done either when a change is made in the policy table or at query time for more dynamic scenarios (our algorithm is efficient enough for dynamic scenarios as we show in Section~\ref{sect:exp}). We later explain how \oursystem can be implemented in existing DBMSs and how it selects an appropriate strategy depending on the query and the set of policies.

\section{Creating Guarded Expressions}
\label{sect:guardSel}

Our goal is to translate a policy expression  $\vPolicyExpression{\vPolicySet{}}= \vObjectConditions{1}{} \lor \cdots \lor \vObjectConditions{\vSetCardinality{\vPolicySet{}}}{}$ into a guarded policy expression $\vPolicyGuardedExpression{\vPolicySet{}} = \vGuard{}{1} \lor \cdots \lor \vGuard{}{n}$ such that the cost of evaluating $\vPolicyGuardedExpression{\vPolicySet{}}$ given $\vDatabase$ and $\vIndexSet{}$\footnote{For our purposes we will assume that the set of available indexes is known.} is minimized
\begin{equation}\label{eq:minCost}
\min cost (\vPolicyGuardedExpression{\vPolicySet{}},\vGuard{}{}) = \min \sum_{\vGuard{}{i} \in \vGuard{}{}} cost(\vGuard{}{i})   
\end{equation}

\noindent where $\vGuard{}{}$ is the set of all the guarded expressions in $\vPolicyGuardedExpression{\vPolicySet{}}$. A guarded expression $\vGuard{}{i}$ corresponds to $\vGuard{}{i}=\vObjectCondition{i}{g} \land \vPolicySet{\vGuard{}{i}}$ where $\vObjectCondition{i}{g}$ is an object condition on an indexed attribute. To model $cost(\vGuard{}{i})$ let us define first the cost of evaluating a tuple against a set of policies as
\begin{equation}\label{eq:costEvalTuplePolicies}
cost(eval(\vPolicyExpression{\vPolicySet{\vGuard{}{i}}}, \vTuple{}{t})) = \vShortCircuit.\vSetCardinality{\vPolicySet{\vGuard{}{i}}}.\vEvalCost
\end{equation}

\noindent where $\vShortCircuit$ represents the average number of policies in $\vPolicySet{\vGuard{}{i}}$ that the tuple $\vTuple{}{t}$ is checked against before it satisfies one (as the policies in $\vPolicyExpression{\vPolicySet{\vGuard{}{i}}}$ form a disjunctive expression\footnote{We assume that the execution of such disjunctive expression stops with the first policy condition evaluating to true and skipping the rest of the policy conditions.}), and $\vEvalCost$ represents the average cost of evaluating $\vTuple{}{t}$ against the set of object conditions for a policy $\vPolicy{}{l} \in \vPolicySet{\vGuard{}{i}}$ (i.e., $\vObjectConditions{l}{}$). The values of $\vShortCircuit$ and $\vEvalCost$ are determined experimentally using a set of sample policies and tuples. Hence, we model $cost(\vGuard{}{i})$ as
\begin{equation}\label{eq:costPartition}
cost(\vGuard{}{i}) = \vSelectivityMethod{\vObjectCondition{i}{g}}.(\vReadCost + cost(eval(\vPolicyExpression{\vPolicySet{\vGuard{}{i}}}, \vTuple{}{t})))
\end{equation}

\noindent where $\vSelectivityMethod{\vObjectCondition{i}{g}}$ denotes the estimated cardinality\footnote{Estimated using histograms maintained by the database.} of the guard $\vObjectCondition{i}{g}$ and $\vReadCost$ represents the cost of reading a tuple from the disk (the value of $\vReadCost$ is also obtained experimentally). Given this cost model, the number of policies in  $\vPolicySet{\vGuard{}{i}}$ and selectivity of $\vObjectCondition{i}{g}$ contribute to most of the cost when evaluating $\vGuard{}{i}$.

The first step in determining $\vPolicyGuardedExpression{\vPolicySet{}}$ is to generate all the {\em candidate guards} ($\vCandidateGuardSet{}{}$), given the object conditions from $\vPolicySet{}{}$, which satisfy the properties of guards as explained in Section~\ref{sect:ourApproach}. Different choices may exist for the same policy given $\vIndexSet{}$; the second step is to select a subset of guards from $\vCandidateGuardSet{}{}$ with the goal of minimizing the evaluation cost of $\vPolicyGuardedExpression{\vPolicySet{}}$.

\subsection{Generating Candidate Guards}

Each policy $\vPolicy{}{l} \in \vPolicySet{\vQuery{}{j}}$ is guaranteed to have at least one object condition (i.e., $\vObjectCondition{l}{owner}$), that trivially satisfies the properties of a guard as 1) $\vObjectCondition{l}{owner}.val$ is a constant and $\vObjectCondition{l}{owner}.attr \in \vIndexSet{}$; and 2) $\vObjectCondition{l}{owner} \in \vObjectConditions{l}{}$. Therefore, we first include all the $\vObjectCondition{l}{k}$ in $\vCandidateGuardSet{}{}$. Similarly, any $\vObjectCondition{l}{c}$ on an indexed attribute with a constant value, belonging to any policy $\vPolicy{}{l}$, can be added to the candidate set $\vCandidateGuardSet{}{}$. However, if only those were to be used as guards, then the size of their corresponding policy partitions $\vPolicySet{\vGuard{}{i}}$ might be small as only policies defined by the same person or policies with the exact same object condition (including, attribute, value, and operation) would be grouped by such a guard.
Exploiting the property that different policies might have common object conditions reduces the number of $\vGuard{}{i}$ and increases the size of their policy partitions $\vPolicySet{\vGuard{}{i}}$ thus decreasing the potential number of evaluations. 
Hence, we create additional candidate guards by {\em merging} range object conditions of different policies on the same attribute, but with different constant values (e.g., if the conditions of two policies on attribute $a$ are $3<a<10$ and $4<a<15$, respectively, the condition $3<a<15$ could be created as a guard, by merging the two conditions, to group both policies in its policy partition). The following theorem limits object conditions that should be considered for this merge based on their overlap.

\begin{theorem}{\label{the:mergeTheorem}} 
Given two candidate guards $\vObjectCondition{x}{c}$ = ($attr^{x}_{1}, op^{x}_{1}$, $val^{x}_{1}, op^{x}_{2}, val^{x}_{2}$) $\in \vObjectConditions{x}{}$, $\vObjectCondition{y}{c} = (attr^{y}_{1}, op^{y}_{1}, val^{y}_{1}, op^{y}_{2}, val^{y}_{2}) \in \vObjectConditions{y}{}$ such that $attr^{x}_{1} = attr^{y}_{1}$ and $attr^{x}_{1} \in \vIndexSet{}$, it is not beneficial to generate a guard by merging them as $\vObjectCondition{x \oplus y}{c}$ = ($attr^{x}_{1}$, $op^{x}_{1}$, $val^{x \oplus y}_1, op^{y}_{2}, val^{x \oplus y}_2$) with $val^{x \oplus y}_1=min(val^{x}_{1},val^{y}_{1})$ and $val^{x \oplus y}_2=max(val^{x}_{2},val^{y}_{2})$ iff $[val^{x}_{1}, val^{x}_{2}] \cap [val^{y}_{1}, val^{y}_{2}]=\phi$.
\end{theorem} 

\textit{Proof:} By Equation~\ref{eq:costPartition}, and considering a guarded expression that contains only a single policy $\vPolicy{}{l}$, the evaluation cost using $\vObjectCondition{l}{c}$ as guard is given by
\begin{equation}\label{eq:costPolicy}
\vCostMethod{\vPolicy{}{l}} = \vSelectivityMethod{\vObjectCondition{l}{c}}.(\vReadCost + \vEvalCost)
\end{equation}

Given two policies $\vPolicy{}{x}$ and $\vPolicy{}{y}$ with candidate guards $\vObjectCondition{x}{c}$ and $\vObjectCondition{y}{c}$\footnote{For simplification of notation in this proof we use $\vObjectCondition{x/y}{c}$ to denote the values in the range $[val^{x/y}_{1}, val^{x/y}_2]$.} such that $\vObjectCondition{x}{c} \cap \vObjectCondition{y}{c} = \emptyset$, it is trivial to see that the cost of evaluating their merge is always greater than evaluating them separately. W.l.o.g., let us consider that $min(val^{x}_{1},val^{y}_{1})=val^{x}_{1}$ and $max(val^{x}_{2},val^{y}_{2})=val^{y}_{2}$ hence the evaluation cost if they were to be merged would be
\begin{multline}\label{eq:costMergeEmptyIntersection}
\vCostMethod{\vPolicy{}{x}\oplus\vPolicy{}{y}} = \vSelectivityMethod{\vObjectCondition{x \oplus y}{c}}.(\vReadCost + \vEvalCost)= \\
(\vSelectivityMethod{\vObjectCondition{x}{c}}+\vSelectivityMethod{\vObjectCondition{y}{c}}).(\vReadCost + \vEvalCost)+\vSelectivityMethod{\vObjectCondition{e}{c}}.(\vReadCost + 2.\vEvalCost)
\end{multline}

\noindent where $\vObjectCondition{e}{c}= (attr^{x}_{1}, op^{x}_{1}, val^{x}_{1}, op^{y}_{2}, val^{y}_{2})$ and hence $\vSelectivityMethod{\vObjectCondition{e}{c}}>=0$, which makes $\vCostMethod{\vPolicy{}{x}\oplus\vPolicy{}{y}}>=\vCostMethod{\vPolicy{}{x}}+\vCostMethod{\vPolicy{}{y}}$. \qed

For situations where $[val^{x}_{1}, val^{x}_{2}] \cap [val^{y}_{1}, val^{y}_{2}] \neq \phi$ we can derive the condition that will make merging beneficial. As previously, let us consider w.l.o.g. that $min(val^{x}_{1},val^{y}_{1})=val^{x}_{1}$ and $max(val^{x}_{2},val^{y}_{2})=val^{y}_{2}$. If the candidate guards were to be merged the new cost of evaluation would be given by $\vCostMethod{\vPolicy{}{x}\oplus\vPolicy{}{y}} = \vSelectivityMethod{\vObjectCondition{x}{c} \cup \vObjectCondition{y}{c}}.(\vReadCost + 2.\vEvalCost)$ which, applying the inclusion-exclusion principle, becomes
\begin{multline}\label{eq:costMergeNotEmptyIntersection}
    \vCostMethod{\vPolicy{}{x}\oplus\vPolicy{}{y}} = (\vSelectivityMethod{\vObjectCondition{x}{c}} + \vSelectivityMethod{\vObjectCondition{y}{c}} - \\
    \vSelectivityMethod{\vObjectCondition{x}{c} \cap \vObjectCondition{y}{c}}).(\vReadCost + 2.\vEvalCost)
\end{multline}

Given that merging will be beneficial if $\vCostMethod{\vPolicy{}{x}\oplus\vPolicy{}{y}}<\vCostMethod{\vPolicy{}{x}}+\vCostMethod{\vPolicy{}{y}}$ and by Equations~\ref{eq:costPolicy} and~\ref{eq:costMergeNotEmptyIntersection} we have
\begin{equation}
    \begin{split}
    &  (\vSelectivityMethod{\vObjectCondition{x}{c}} + \vSelectivityMethod{\vObjectCondition{y}{c}} - \vSelectivityMethod{\vObjectCondition{x}{c} \cap \vObjectCondition{y}{c}}).(\vReadCost + 2.\vEvalCost) < \\
    & \vSelectivityMethod{\vObjectCondition{x}{c}}.(\vReadCost + \vEvalCost) + \vSelectivityMethod{\vObjectCondition{y}{c}}.(\vReadCost + \vEvalCost) \\
    & \vSelectivityMethod{\vObjectCondition{x}{c}}.\vEvalCost + \vSelectivityMethod{\vObjectCondition{y}{c}}.\vEvalCost-\vSelectivityMethod{\vObjectCondition{x}{c} \cap \vObjectCondition{y}{c}}(\vReadCost + 2.\vEvalCost)<0
    \end{split}
\end{equation}

\noindent Using inclusion exclusion principle if follows that
\begin{equation}{\label{eq:overlapCon}}
    \dfrac{\vSelectivityMethod{\vObjectCondition{x}{c} \cap \vObjectCondition{y}{c}}}{\vSelectivityMethod{\vObjectCondition{x}{c} \cup \vObjectCondition{y}{c}}} > \dfrac{\vEvalCost}{\vReadCost + \vEvalCost}
\end{equation}

\noindent which is the condition to be checked to merge those two overlapping candidates. Equation~\ref{eq:overlapCon} is checked by the function $\vMergeMethod{\vObjectCondition{x}{c}}{\vObjectCondition{y}{c}}$ which returns $\vObjectCondition{x \oplus y}{c}$ if merging $\vObjectCondition{x}{c}$ and $\vObjectCondition{y}{c}$ is beneficial and $\phi$ otherwise. 
$\vObjectCondition{x \oplus y}{c}$ is added to $\vCandidateGuardSet{}{}$ and we maintain in a mapping structure that both $\vPolicy{}{x}$ and $\vPolicy{}{y}$ are relevant to that candidate guard (this information will be used in the second step). 
Given Theorem~\ref{the:mergeTheorem}, $\vMergeMethod{\vObjectCondition{x}{c}}{\vObjectCondition{y}{c}}$ is computed only if $\vSetCardinality{\vObjectCondition{x}{c} \cap \vObjectCondition{y}{c}} \neq \phi$, we first order the candidate guards by their left range value in ascending order.
The number of checks to be done could still be high as a candidate guard could potentially merge with another transitively. For example, given a situation where $\vObjectCondition{x}{c} \cap \vObjectCondition{y}{c} \neq \phi$, $\vObjectCondition{y}{c} \cap \vObjectCondition{z}{c} \neq \phi$, and $\vObjectCondition{x}{c} \cap \vObjectCondition{z}{c} = \phi$, which might make $\vMergeMethod{\vObjectCondition{x}{c}}{\vObjectCondition{y}{c}} \neq \phi$, $\vMergeMethod{\vObjectCondition{y}{c}}{\vObjectCondition{z}{c}} \neq \phi$, and $\vMergeMethod{\vObjectCondition{x}{c}}{\vObjectCondition{z}{c}} = \phi$, it could be possible the transitive merge of $\vObjectCondition{x}{c}$ with $\vObjectCondition{y \oplus z}{c}$ is beneficial (i.e., $\vMergeMethod{\vObjectCondition{x}{c}}{\vObjectCondition{y \oplus z}{c}} = \phi$). 
We present a condition to limit the number of checks to be performed due to such transitive overlaps for a given a $\vCandidateGuardSet{}{}$ with candidate guards sorted in the ascending order of their left range values.

First, we show as a consequence of Theorem~\ref{the:mergeTheorem} that transitive merges will not be useful under the following condition.

\begin{corollary}{\label{theorem:pruningCon}}
Given two candidate guards $\vObjectCondition{x}{c}$ and $\vObjectCondition{y}{c}$, such that $\vObjectCondition{x}{c} \cap \vObjectCondition{y}{c} \neq \phi$ and whose merging is not beneficial (i.e., $\vMergeMethod{\vObjectCondition{x}{c}}{\vObjectCondition{y}{c}} = \phi$), and given another candidate guard $\vObjectCondition{y \oplus z}{c}$, generated after merging $\vObjectCondition{y}{c}$ and $\vObjectCondition{z}{c}$, the transitive merge of $\vObjectCondition{x}{c}$ and $\vObjectCondition{y \oplus z}{c}$ will not be beneficial (i.e., $\vMergeMethod{\vObjectCondition{x}{c}}{\vObjectCondition{y \oplus z}{c}} = \phi$) if $\vObjectCondition{x}{c} \cap \vObjectCondition{z}{c} = \phi$.
\end{corollary}

Let us consider $\vObjectCondition{x}{c} \cap \vObjectCondition{z}{c} = \phi$.
By Equation~\ref{eq:costPolicy} and Equation~\ref{eq:costMergeNotEmptyIntersection}, we calculate the cost of such a merge by
\begin{multline}
\vCostMethod{\vPolicy{}{x}\oplus(\vPolicy{}{y\oplus z})} =  (\vSelectivityMethod{\vObjectCondition{x}{c}}+\vSelectivityMethod{\vObjectCondition{y}{c}} + \vSelectivityMethod{\vObjectCondition{z}{c}} + \\ \vSelectivityMethod{\vObjectCondition{x}{c} \cap \vObjectCondition{y}{c}} + \vSelectivityMethod{\vObjectCondition{y}{c} \cap \vObjectCondition{z}{c}}).(\vReadCost + 3.\vEvalCost) 
\end{multline}

\noindent which makes $\vCostMethod{\vPolicy{}{x}\oplus(\vPolicy{}{y}\oplus\vPolicy{}{z})}>\vCostMethod{\vPolicy{}{x}}+\vCostMethod{\vPolicy{}{y}}+\vCostMethod{\vPolicy{}{z}}$ and hence $\vMergeMethod{\vObjectCondition{x}{c}}{\vObjectCondition{y \oplus z}{c}} = \phi$.

In addition, in the situation described in Corollary~\ref{theorem:pruningCon}, we can show that there is no need to merge $\vObjectCondition{x}{c}$ with any other candidate following $\vObjectCondition{z}{c}$.

\begin{corollary}
Given the situation explained in Corollary~\ref{theorem:pruningCon}, let us define  $\hat{\vCandidateGuardSet{}{}}=\vCandidateGuardSet{}{}\setminus\{\vObjectCondition{x}{c},\vObjectCondition{y}{c},\vObjectCondition{z}{c}\}$. For any $\vObjectCondition{w}{c} \in \hat{\vCandidateGuardSet{}{}}$, the transitive merge with $\vObjectCondition{x}{c}$ is not beneficial (i.e., $\vMergeMethod{\vObjectCondition{x}{c}}{\vObjectCondition{y \oplus z \oplus w}{c}} = \phi \; \forall \; \vObjectCondition{w}{c} \in \hat{\vCandidateGuardSet{}{}}$).
\end{corollary}

As the candidate guards are sorted by their left ranges and $\vObjectCondition{x}{c} \cap \vObjectCondition{z}{c} = \phi$, we also have $\vObjectCondition{x}{c} \cap \vObjectCondition{w}{c} = \phi$. Therefore, as shown in Corollary~\ref{theorem:pruningCon}, the transitive merge with $\vObjectCondition{x}{c}$, $\vObjectCondition{y}{c}$, $\vObjectCondition{z}{c}$, and $\vObjectCondition{w}{c}$ will not be beneficial ($\vMergeMethod{\vObjectCondition{x}{c}}{\vObjectCondition{y \oplus z \oplus w}{c}} = \phi$).

To summarize, the steps for generating $\vCandidateGuardSet{}{}$ from a set of policies $\vPolicySet{}$ are then as follows: 1) For all $\vPolicy{l}{} \in \vPolicySet{}$ collect object conditions that satisfy guard properties by their attribute; 2) For each such collection sort range object conditions by their left range; 3) For the first candidate guard ($\vObjectCondition{1}{c}$), verify whether the next candidate guard ($\vObjectCondition{2}{c}$) is such that $\vMergeMethod{\vObjectCondition{1}{c}}{\vObjectCondition{2}{c}} \neq \phi$. In that case, merge both candidate guards to generate $\vObjectCondition{1 \oplus 2}{c}$ which is added to $\vCandidateGuardSet{}{}$ ($\vPolicy{}{1}$ and $\vPolicy{}{2}$ get associated to the new merged candidate). Otherwise, if $\vMergeMethod{\vObjectCondition{1}{c}}{\vObjectCondition{2}{c}} = \phi$, then we check $\vObjectCondition{1}{c}$ with the following candidate guards until the condition in Corollary~\ref{theorem:pruningCon} is satisfied and move to the next candidate guard when it does and repeat the process.

\subsection{Selecting Cost Optimal Guards}

We next select the subset of guards  $\vGuard{}{} \in \vCandidateGuardSet{}{}$ that minimizes the cost according to Equation~\ref{eq:minCost}. The guard selection problem can be formally stated as 
\begin{multline}\label{eq:minimizationProblem}
\min_{\vGuard{}{} \subseteq \vCandidateGuardSet{}{}} \vCostMethod{\vGuard{}{}}
    = \\ \sum_{\vGuard{}{i}\in \vGuard{}{}} \vCostMethod{\vGuard{}{i}} \; \forall \; \vPolicy{}{i} \in \vPolicySet{} \;
    \exists \; \vGuard{}{i} \in \vGuard{}{} \mid \vPolicy{}{i} \in \vPolicySet{\vGuard{}{i}}    
\end{multline}

The problem of selecting $\vGuard{}{}$ from $\vCandidateGuardSet{}{}$ such that every policy in $\vPolicySet{}$ is covered exactly once (as this would limit the extra checkings) can be shown to be NP-hard by reducing weighted Set-Cover problem to it. In the weighted Set-Cover problem, we have a set of elements $E = {e_1, \cdots, e_n}$ and a set of subsets over $E$ denoted by $S = {S_1, \cdots, S_m}$ with each set $S_i \in S$ having a weight $w_i$ associated with it. The goal of set cover problem is to select $\min_{\hat{S} \subseteq S} \sum{S_i.w_i} \mid S_i \in \hat{S}$  and $E = \bigcup_{S_i \in \hat{S}} S_i$. From our guard selection problem we have $E$ and $S$ equivalent to $\vPolicySet{}$ and $\vCandidateGuardSet{}{}$ respectively. We assign $e_i$ to $S_i$ when the corresponding $\vPolicy{}{i}$ is assigned to $\vGuard{}{i}$.
The weight function $w_i$ set to $\vCostMethod{\vGuard{}{i}}$ where the evaluation cost of $\vPolicyGuardedExpression{\vGuard{}{i}}$ is set to zero. So we have $w_i = \vSetCardinality{\vObjectConditions{i}{c}}.\vReadCost$. If a polynomial time algorithm existed to solve this problem, then it would solve set-cover problem too.

The evaluation cost of a policy depends on the guard it is assigned to. We define a utility heuristic\footnote{Similar to the one used by~\cite{hellerstein1998optimization} for optimizing queries with expensive predicates.} which ranks the guards by their benefit per unit read cost. Without a guard, $\vPolicyExpression{\vPolicySet{}}$ will be evaluated by a linear scan followed by the checking of $\vPolicyExpression{\vPolicySet{}}$ as filter on top of $\vTuple{}{t} \in \vTupleSet{\vQuery{}{i}}$. The guard $\vObjectCondition{i}{c} \in \vGuard{}{}$ reduces the number of tuples that have to be checked against each $\vPolicyExpression{\vPolicySet{\vGuard{}{i}}}$. The benefit of a guard captures this difference by $\vBenefitMethod{\vGuard{}{i}} = \vEvalCost.\vSetCardinality{\vPolicySet{\vGuard{}{i}}}.(\vSetCardinality{\vRelation{}{i}} - \vSelectivityMethod{\vObjectCondition{i}{c}})$. Using this benefit method, and the read cost of evaluating $\vGuard{}{i}$ defined earlier, we define the utility of $\vGuard{}{i}$ as $\vUtilityMethod{\vGuard{}{i}}= \frac{\vBenefitMethod{\vGuard{}{i}}}{\vReadCostMethod{\vGuard{}{i}}}$.

Algorithm~\ref{alg:guardSelection} uses this heuristic to select the best possible guards to minimize the cost of policy evaluation. First, it iterates over $\vCandidateGuardSet{}{}$ and stores each guarded expression $\vGuard{}{i} \in \vCandidateGuardSet{}{}$ (comprised of a guard $\vObjectCondition{g}{i}$ and a policy partition $\vPolicySet{\vGuard{}{i}}$) in a priority queue in descending order of their utility. 
Next, the priority queue is polled for the $\vGuard{}{i}$ with the highest utility. If $\vPolicySet{\vGuard{}{i}}$ intersects with another $\vPolicySet{\vGuard{}{j}} \in \vCandidateGuardSet{}{}$, $\vPolicySet{\vGuard{}{j}}$ is updated to remove the intersection of policies and $\vUtilityMethod{\vGuard{}{j}}$ is recomputed after which the new $\vGuard{}{j}$ is reinserted into priority queue in the order of its utility. The result is thus the subset of candidates guards $\vGuard{}{}$ that maximizes the benefit and covers all the policies in $\vPolicySet{}$, that is, minimizes  $cost(\vPolicyGuardedExpression{\vPolicySet{}},\vGuard{}{})$ in Equation~\ref{eq:minimizationProblem}.

\begin{algorithm}
\scriptsize
\caption{Selection of guards}
\begin{algorithmic}[1]
\Function{GuardSelection}{$\vCandidateGuardSet{}{}$}
    \For{ i in 1 $\cdots$ $\vSetCardinality{\vCandidateGuardSet{}{}}$ }
        \State C[i] = \Call{cost}{$\vGuard{}{i}$}; U[i] = \Call{utility}{$\vGuard{}{i}$}
    \EndFor
    \State $Q \gets \phi$
    \For{ i in 1 $\cdots$ $\vSetCardinality{\vCandidateGuardSet{}{}}$ }
        \State \Call{PriorityInsert}{$Q$,$\vGuard{}{i}$,$U[i]$}
    \EndFor
    \While{$Q$ is not empty}
        \State $\vGuard{}{max} =$ \Call{Extract-Maximum}{$Q$}; $\vGuard{}{} \gets \vGuard{}{max}$
        \ForEach{$\vGuard{}{i}$ in $Q$}
            \If{$\vPolicySet{\vGuard{}{i}} \cap \vPolicySet{\vGuard{}{max}} \neq \phi$}
                \State $\vPolicySet{\vGuard{}{i}} = \vPolicySet{\vGuard{}{i}} \setminus \vPolicySet{\vGuard{}{max}}$; \Call{Remove}{$Q, \vGuard{}{i}$}
                \If{$\vPolicySet{\vGuard{}{i}} \neq \phi$} 
                    \State B = \Call{benefit}{$\vGuard{}{i}$}; U[i] = $\dfrac{B}{C[i]}$
                    \State \Call{PriorityInsert}{$Q$,$\vGuard{}{i}$,$U[i]$}
                \EndIf
               
            \EndIf
        \EndFor
    \EndWhile
    \Return $\vGuard{}{}$
\EndFunction
\end{algorithmic}
\label{alg:guardSelection}
\end{algorithm}

\vspace{-0.6cm}
\section{Implementing \oursystem}
\label{sect:implementingSieve}

\oursystem is a general-purpose middleware that intercepts queries posed to a database, optimally rewrites them, and submits the queries to the underlying database on which it is layered for execution. SIEVE rewrites queries such that the rewritten queries can be executed efficiently to produce query results that are compliant with the policies. SIEVE's rewriting is based on: (a) decreasing the policies that have to be check per tuple and (b) reducing the number of tuples that have to be checked against policy expressions. In implementing this, \oursystem exploits the extensibility options of databases such as support for UDFs and index usage hints. The implementation of \oursystem with connectors for both MySQL and PostgreSQL is available at \url{https://github.com/primalpop/sieve}.

\ifextended
\subsection{Persistence of Policies and Guards}

To store policies associated with all the relations in the database, \oursystem uses two additional relations, the policy table (referred to as $\vRelation{}{P}$), which stores the set of policies, and the object conditions table (referred to as $\vRelation{}{OC}$), which stores conditions associated with the policies. The structure of $\vRelation{}{P}$ corresponds to |$\langle$id, owner, querier, associated-table, purpose, action, ts-inserted-at$\rangle$|, where |associated-table| is the relation $\vRelation{}{i}$ for which the policy is defined and |ts-inserted-at| is the timestamp at policy insertion. The schema of $\vRelation{}{OC}$ corresponds to |$\langle$policy-id, attr, op, val$\rangle$| where |policy-id| is a foreign key to $\vRelation{}{P}$ and the rest of attributes represent the condition $\vObjectCondition{l}{c}$=\vTupleExpression{$attr$}{$op$}{$val$}. 
We emphasize that the value |val| in $\vRelation{}{OC}$ might correspond to a complex SQL condition in case of nested policies. For instance, the two sample policies defined in Section~\ref{sect:modeling} regulate access to student connectivity data for Prof. Smith; they are persisted as tuples |$\langle$1, John, Prof.Smith, WiFiDataset, Attendance Control, Allow, 2020-01-01 00:00:01$\rangle$| and |$\langle$2, John, Prof.Smith, WiFiDataset, Attendance Control, Allow, 2020-01-01 00:00:01$\rangle$| in $\vRelation{}{P}$ and with the tuples |$\langle$1, 1, wifiAP, $=$, 1200$\rangle$|, |$\langle$2, 1, ts-time, $\geq$, 09:00$\rangle$|, |$\langle$3, 1, ts-time, $\leq$, 10:00$\rangle$|, |$\langle$4, 2, wifiAP, $=$, SELECT W2.wifiAP FROM WiFiDataset AS W2 WHERE W2.owner = "Prof.Smith" and W2.ts-time = W.ts-time$\rangle$| in $\vRelation{}{OC}$.

A guarded policy expression $\vPolicyGuardedExpression{\vPolicySet{}}$ generated, per user and purpose, is stored in $\vRelation{}{GE}$ with the schema |$\langle$id, querier, associated-table, purpose, action, outdated, ts-inserted-at$\rangle$|. Guarded policy expressions are not continuously updated based on incoming policies as this would be unnecessary if their specific queriers do not pose any query. We use the |outdated| attribute, which is a boolean flag, to describe whether the guarded expression includes all the policies belonging to the querier. If at query time, the |outdated| attribute associated to the guarded policy expression for the specific querier/purpose (as specified in the query metadata $\vQueryMetadata{i}{querier}, \vQueryMetadata{i}{purpose}$) is found to be true, then that guarded policy expression is regenerated. After the guarded expression is regenerated for a querier, it is stored in the table with |outdated| set to false. 
Guard regeneration comes with an overhead. However, in our experience, the corresponding overhead is much less than the execution cost of queries. As a result, we generate guards during query execution using triggers in case the current guards are outdated. 
Guarded expressions $\vGuard{}{i}$ associated with a guarded policy expression $\vPolicyGuardedExpression{}$ are stored in two relations:  $\vRelation{}{GG}$=|$\langle$id, guard-expression-id, attr, op, val$\rangle$| to store the guard (i.e., $\vObjectCondition{i}{g}$=\vTupleExpression{$attr$}{$op$}{$val$}) and $\vRelation{}{GP}$=|$\langle$guard-id, policy-id$\rangle$| to store the policy partition (i.e., $\vPolicySet{\vGuard{}{i}}$).
\fi

\subsection{Implementing Operator $\Delta$}
\label{sect:UDF}

We implement the policy evaluation operation $\Delta$ (see Section~\ref{sect:ourApproach}) by User Defined Functions (UDFs) on top of a DBMS. 
\ifextended
Consider a set of policies $\vPolicySet{}$ and the query metadata $\vQueryMetadata{i}{}$ and a tuple $\vTuple{}{t}$ belonging to relation $\vRelation{}{j}$. $\Delta(\vPolicySet{}, \vQueryMetadata{i}{}, \vTuple{}{t})$ is implemented as the following UDF
\begin{lstlisting}
CREATE FUNCTION delta($[policy]$, $querier$, $purpose$, $[attrs]$) 
{BEGIN
 Cursor c = 
  SELECT $\vRelation{}{OC}.attr$ as attr, $\vRelation{}{OC}.op$ as op, $\vRelation{}{OC}.val$ as val
  FROM $\vRelation{}{P}, \vRelation{}{OC}$ 
  WHERE $\vRelation{}{P}.querier = querier$ AND $\vRelation{}{P}.purpose = purpose$ AND $\vRelation{}{P}.id$ IN $[policy]$ AND $\vRelation{}{P}.owner = [attrs].owner$ AND $\vRelation{}{P}.id = \vRelation{}{OC}.policy-id$
  LET satisfied_flag = true
  READ UNTIL c.isNext() = false:
     FETCH c INTO p_attr, p_op, p_val
      FOR each t_attr in $[attrs]$
        IF t_attr = p_attr THEN
          satisfied_flag = satisfied_flag AND /*Check whether t_val satisfies p_op p_val*/
  return satisfied_flag
END}
\end{lstlisting}

The UDF above performs two operations: 1) It takes a set of policies and retrieves a subset $\hat{\vPolicySet{}}$ which contains the relevant policies to be evaluated based on the query metadata $\vQueryMetadata{i}{}$ and the tuple $\vTuple{}{t}$; 2) It performs the evaluation of each policy $\vPolicy{}{i} \in \hat{\vPolicySet{}}$ on $\vTuple{}{t}$. 
\else
Given a set of policies $\vPolicySet{}$, the query metadata $\vQueryMetadata{i}{}$, and a tuple $\vTuple{}{t}$ belonging to relation $\vRelation{}{j}$. $\Delta(\vPolicySet{}, \vQueryMetadata{i}{}, \vTuple{}{t})$ is implemented as a UDF which performs two operations: 1) It takes a set of policies and retrieves a subset $\hat{\vPolicySet{}}$ which contains the relevant policies to be evaluated based on the query metadata $\vQueryMetadata{i}{}$ and the tuple $\vTuple{}{t}$. This way, it retrieves the policies that are defined by the owner of the tuple for the specific querier and her purpose. 2) It performs the evaluation of each policy $\vPolicy{}{i} \in \hat{\vPolicySet{}}$ on $\vTuple{}{t}$. 
\fi

\subsection{Implementing Policy Guarded Expression}
\label{sect:implementingGuards}
 
Our goal is to evaluate policies for query $\vQuery{}{i}$ by replacing any relation $\vRelation{}{j} \in \vQuery{}{i}$ by a projection of $\vRelation{}{j}$ that satisfies the guarded policy expression $\vPolicyGuardedExpression{\vPolicySet{\vRelation{}{j}}}$ where $\vPolicySet{\vRelation{}{j}}$ is the set of policies defined for the specific querier, purpose, and relation. To this end, we first use the |WITH| clause for each relation $\vRelation{}{j} \in \vQuery{}{i}$ that selects tuples in $\vRelation{}{j}$  satisfying the guarded policy expression\footnote{Using the above strategy the policy check needs to be only done once in the WITH clause even if the relation appears multiple times in the query.}. The rewritten query replaces every occurrence of $\vRelation{}{j}$ with the corresponding $\hat{\vRelation{}{j}}$. 

\begin{lstlisting}
  WITH $\hat{\vRelation{}{j}}$ AS (
    SELECT * FROM $\vRelation{}{j}$ WHERE $\vGuard{}{1}$ OR $\vGuard{}{2}$ OR $\cdots$ OR $\vGuard{}{n}$)
\end{lstlisting}

\oursystem utilizes extensibility features (e.g., index usage hints\footnote{\url{https://dev.mysql.com/doc/refman/8.0/en/index-hints.html}}, optimizer explain\footnote{\url{https://www.postgresql.org/docs/13/sql-explain.html}}, UDFs) offered by DBMSs that allows it to suggest index plans to the underlying optimizer. Since such features vary across DBMSs, guiding optimizers requires a platform dependent connector that can rewrite the query appropriately. In systems such as MySQL, Oracle, DB2, and SQL Server that support index usage hints, \oursystem can rewrite the query to explicitly force indexes on guards. For example, in MySQL using |FORCE INDEX| hints, which tell the optimizer that a table scan is very expensive and should only be used if the DBMS cannot use the suggested index to find rows in the table, the rewritten query will be as follows:

\begin{lstlisting}
  WITH $\hat{\vRelation{}{j}}$ AS (
    SELECT * FROM $\vRelation{}{j}$ [FORCE INDEX ($\vObjectCondition{1}{g}$)] WHERE $\vGuard{}{1}$ UNION 
    SELECT * FROM $\vRelation{}{j}$ [FORCE INDEX ($\vObjectCondition{2}{g}$)] WHERE $\vGuard{}{2}$ UNION$\cdots$ 
    SELECT * FROM $\vRelation{}{j}$ [FORCE INDEX ($\vObjectCondition{n}{g}$)] WHERE $\vGuard{}{n}$)
\end{lstlisting}

Some systems, like PostgreSQL, do not support index hints explicitly. In such cases, \oursystem still does the above rewrite but depends upon the underlying optimizer to select appropriate indexes.

\subsection{Combining $\Delta$ with Guards}
\label{sect:selectingGuardStrategy}

Depending upon the number of policies in the associated guard partition (i.e., $\vSetCardinality{\vPolicySet{\vGuard{}{i}}}$), we could rewrite the policy partition part using the $\Delta$ operator as $\Delta(\vPolicySet{\vGuard{}{i}}, \vQueryMetadata{i}{}, \vTuple{}{t})$ if it reduces the execution cost,  instead of checking the polices inline as shown in Section~\ref{sect:implementingGuards}. 
The $\Delta$ operator has an associated cost due to the invocation and execution of a UDF. For each guarded expression $\vGuard{}{i}$ in a guarded policy expression $\vPolicyGuardedExpression{}$ for a relation $\vRelation{}{i}$ and a specific querier and purpose, we check the overhead of using the $\Delta$ operator (to which we will refer to as $Guard\&\Delta$) versus not using it ($Guard\&Inlining$) and use $\Delta$ if $\vCostMethod{Guard\&\Delta}<\vCostMethod{Guard\&Inlining}$.

We model the cost of each strategy by computing the cost of evaluating policies per tuple since the number of tuples to check are the same in both cases. As modeled in Equation~\ref{eq:costEvalTuplePolicies},  $\vCostMethod{Guard\&Inlining}=\vShortCircuit.\vSetCardinality{\vPolicySet{\vGuard{}{i}}}.\vEvalCost$ where the values of $\vShortCircuit$, the percentage of policies that have to be checked before one returns true, and $\vEvalCost$, the cost of evaluating a policy against a single tuple, are obtained experimentally. We compute $\vShortCircuit$ by executing a query which counts the number of policy checks done over $\vPolicySet{\vGuard{}{i}}$ before a tuple either satisfies one of the policies or is discarded (does not satisfy any policy) and averaging the number of policy checks across all tuples. 
We estimate $\vEvalCost$ by computing the difference of the read cost per tuple without policies (estimated by dividing the time it takes to perform a table scan by the total number of tuples) and the average cost per tuple with policies. The former is estimated by executing a table scan with different number of policies with different selectivities (number of tuples) and averaging the cost per tuple per policy.
$\vCostMethod{Guard\&\Delta} = UDF_{inv} + UDF_{exec}$ where the two factors represent the cost of invocation and execution of the UDF, respectively\footnote{Recent work, such as~\cite{wang2019idea}, shows that in some situations batching of UDF operations might be possible to save the overhead of $UDF_{inv}$ per tuple. While current DBMSs generally lack support for this optimization, our model could be easily adapted to consider such cost amortizations.}. We obtain this cost experimentally by executing $\Delta$ with varying number of tuples (by changing the selection predicate before $\Delta$) and the number of policies to be checked against (by changing the guard associated with invocation of operator).

Most of the terms in both cost models are constants, the term that varies depending on the specific guard is $\vSetCardinality{\vPolicySet{\vGuard{}{i}}}$. Our experiments (see Section~\ref{sect:exp}) indicate that the usage of the $Guard\&\Delta$ strategy is beneficial if $\vSetCardinality{\vPolicySet{\vGuard{}{i}}}>120$.

\subsection{Exploiting Selection Predicates in Queries}
\label{sect:exploitingQuery}

So far, we only considered exploiting guarded policy expressions to optimize the overhead of policy checks while executing a query $\vQuery{}{i}$. 
We could further exploit selection predicates defined over relation $\vRelation{}{j}$ that appears in $\vQuery{}{i}$, especially if such predicates are highly selective, in reducing the cost of policy checking. The rewrite strategy discussed above, that is used to replace $\vRelation{}{j}$ into $\hat{\vRelation{}{j}}$, can be modified to include such selective query predicates in addition to the guarded policy expression.
Such a modification provides the optimizer with a choice on whether to use the index on the guards or to use the query predicate to filter the tuples in the relation on which we apply the policy checks. 

Instead of depending upon the optimizer to choose correctly\footnote{Optimizers might choose suboptimal plans when query predicates are as complex as the guarded policy expressions.}, \oursystem provides a hint to the optimizer based on estimating the cost of different possible execution strategies. In particular, \oursystem considers the following three possibilities: 1) Linear scan of the relation combined with a guarded evaluation of the policy (referred to as $LinearScan$). 2) Index scan based on query predicate followed by the evaluation of the guarded policy expression (referred to as $IndexQuery$). 3) Index scan based on guards followed by evaluation of the policy partitions (referred to as $IndexGuards$). In each of the strategies, guarded expressions are used to generate $\hat{\vRelation{}{j}}$  while access methods used may differ.

To determine cost of each strategy, \oursystem first runs the |EXPLAIN| of query $\vQuery{}{i}$ which returns a high-level view of the query plan including, usually, for each relation in the query the particular access strategy (table scan or a specific index) the optimizer plans to use and estimated selectivity of the predicate on that attribute ($p$). Then, \oursystem estimates an upper bound of the cost for each strategy focusing on the cost of accessing data. $\vCostMethod{IndexGuards}$ is computed as $\sum_{\vGuard{}{i} \in \vGuard{}{}}\vSelectivityMethod{\vGuard{}{i}}.\vReadCost$, where $\vSelectivityMethod{\vGuard{}{i}}$ is the cardinality of the guard $\vObjectCondition{1}{g}$. If the optimizer selects to perform index scan on a query predicate $p$, then $\vCostMethod{IndexQuery} = \vSelectivityMethod{p}.\vReadCost$ otherwise $\vCostMethod{IndexQuery}=\infty$.   
\oursystem chooses between $IndexGuards$ and $IndexQuery$ based on which strategy is less costly. It then compares the better of the two strategies to $LinearScan$ choosing the latter if the random access due to index scan is expected to be more costly than the sequential access of linear of scan.
To implement the selected strategy, \oursystem rewrites the query (including the appropriate |WITH| clause(s) as explained in Section~\ref{sect:implementingGuards}) to append: 
An index hint (e.g., |FORCE INDEX| statement in MySQL) for each guard $\vGuard{}{i}$ to the |FROM| clause within the |WITH| clause as we showed previously (in the case of $IndexGuards$ strategy); or an index hint for the attribute of $p$ (for $IndexQuery$); or a hint to suggest the optimizer to igonore all indexes (e.g., |USE INDEX()| in MySQL) (for $LinearScan$). 

\subsection{Sample Query Rewriting in \oursystem}

Let us consider the query in Section~\ref{sect:caseStudy} to study the tradeoff between student performance and attendance to classes. In that case, \oursystem might rewrite the query as follows depending on the available policies and DBMS:
\begin{lstlisting}
WITH WiFiDatasetPol AS (
 SELECT * FROM WiFiDataset as W FORCE INDEX($\vObjectCondition{1}{g} \cdots \vObjectCondition{n}{g}$)
 WHERE ($\vObjectCondition{1}{g}$ AND W.ts-date between "9/25/19" AND "12/12/19" AND ($\vObjectCondition{1}{1}$ AND $\cdots$ AND $\vObjectCondition{1}{n}$))
        OR $\cdots$ OR 
       ($\vObjectCondition{n}{g}$ AND W.ts-date between "9/25/19" AND "12/12/19" AND delta(32,"Prof.Smith", "Analysis","owner","ts-date", "ts-time", "wifiAP")=true)
) StudentPerf(WifiDatasetPol, Enrollment, Grades)
 \end{lstlisting}
 
As the query has only one table with associated policies in its |FROM| clauses (i.e., |WiFiDataset| table), the rewritten query contains one |WITH| clause, generated as explained in Section~\ref{sect:implementingGuards}. This clause creates |WiFiDatasetPol| which is now used in the original query to replace the |WiFiDataset| table.
The |WITH| clause includes the set of guards generated for the querier (``Prof. Smith'') and his purpose (``Analysis'') given the policies in the database. The query predicate on date (|ts-date between "9/25/19" AND "12/12/19"|) was included along with each guard as outlined in Section~\ref{sect:exploitingQuery}.
As \oursystem selected the $IndexGuards$ strategy, the |WITH| clause forces the usage of guards as indexes (through the |FORCE INDEX| command) as explained in Section~\ref{sect:exploitingQuery}. Finally, for one specific guarded expression ($\vGuard{}{n}$) \oursystem selected the $guard+\Delta$ strategy (see Section~\ref{sect:selectingGuardStrategy}). Hence, its policy partition was replaced by the call to the UDF that implements the $\Delta$ operator. We point out that the implementation of the general UDF presented in Section~\ref{sect:UDF} has been modified slightly to retrieve the policies on the partition of the guard by using the id of the guard (passed as a parameter).

\ifextended
\section{Managing dynamic scenarios}
\label{sect:guardMan}

As mentioned before, the generation of guarded expressions for a set of users can be performed offline. However, in general, the dataset of access control policies defined for a database can change along time (i.e., users add new policies or update existing ones). Hence, \oursystem would need to regenerate guarded expressions to reflect the changes in the policy dataset. The cost associated with guard generation is a function of the number of policies and thus, in situations with very large policy datasets, this cost might not be trivial. Regenerating everytime that a change is made in the policy dataset might not be thus optimal if no queries are executed in between changes. Selecting the frequency of guard regeneration carefully can reduce the total system time. In this section, we first extend the cost model presented earlier to include the query evaluation time. Then, we derive the optimal number of policy insertions before guard regeneration as a function of policy and query rates.

\subsection{Query Evaluation with Guarded Expression}

\noindent The cost of evaluating $\vGuard{}{}$ associated with a $\vUser{}{j}$ is given by 
\begin{equation}{\label{eq:totalGuardCost}}
\vCostMethod{G} = \sum_{\vGuard{}{i} \in \vGuard{}{}} cost(\vGuard{}{i})    
\end{equation}

Given Equation~\ref{eq:costPartition}, and the simplifying assumption that $\vSelectivityMethod{\vObjectCondition{i}{g}}$ is the same for all the guards in $\vGuard{}{}$ and can be represented by $\vSelectivityMethod{\vObjectCondition{}{g}}$, we can express the previous cost as
\begin{align}{\label{eq:totalGuardCostSim}}
\mathclap{\sum_{\vGuard{}{i} \in \vGuard{}{}} cost(\vGuard{}{i})} & \notag \\
&= \sum_{\vGuard{}{i} \in \vGuard{}{}} \vSelectivityMethod{\vObjectCondition{i}{g}}.(\vReadCost+\vEvalCost.\vShortCircuit.\vSetCardinality{\vPolicySet{\vGuard{}{i}}}) \notag \\
&= \vSelectivityMethod{\vObjectCondition{}{g}}.(\vReadCost + \vEvalCost.\vShortCircuit(\vSetCardinality{\vPolicySet{\vGuard{}{1}} + \vPolicySet{\vGuard{}{2}} + \cdots + \vPolicySet{\vGuard{}{m}}}) \notag \\
&= \vSelectivityMethod{\vObjectCondition{}{g}}.(\vReadCost + \vEvalCost.\vShortCircuit.\vSetCardinality{\vPolicySet{n}})
\end{align}

\noindent where $\vSetCardinality{\vPolicySet{\vGuard{}{1}}} + \vSetCardinality{\vPolicySet{\vGuard{}{2}}} + \cdots + \vSetCardinality{\vPolicySet{\vGuard{}{m}}} = \vSetCardinality{\vPolicySet{n}}$ as every policy is exactly covered by one guard. We now define the cost of query evaluation for $\vQuery{}{j}$ (posed by $\vUser{}{j}$) along with $\vGuard{}{}$ (using the $IndexGuards$ approach presented in Section~\ref{sect:implementingGuards}) as
\begin{equation}{\label{eq:guardQuery}}
    \vCostMethod{\vGuard{}{}, \vQuery{}{j}} = \sum_{i=1}^{\vSetCardinality{\vGuard{}{}}} \vCostMethod{\vGuard{}{i}} + \vSelectivityMethod{\vGuard{}{}}.eval(\vPolicyExpression{\vQuery{}{j}}, \vTuple{}{t})
\end{equation}

\noindent where $\vSelectivityMethod{\vGuard{}{}}$ is the cardinality of the guarded expression for $\vUser{}{j}$ (i.e., the number of tuples that satisfy $\vGuard{}{}$ and are then checked against the $\vQuery{}{j}$ posed by $\vUser{}{j}$). We expand this cost using Equation~\ref{eq:totalGuardCostSim} and substitute $\vSelectivityMethod{\vGuard{}{}}$ with $\vSelectivityMethod{\vObjectCondition{}{g}}$ which gives an upper bound of the cost as $\vSelectivityMethod{\vObjectCondition{}{g}} > \vSelectivityMethod{\vGuard{}{}}$.
\begin{equation}{\label{eq:guardQueryFinal}}
    \vCostMethod{\vGuard{}{}, \vQuery{}{j}} = \vSelectivityMethod{\vObjectCondition{}{g}}.(\vReadCost + \vEvalCost.\vShortCircuit.(\vSetCardinality{\vPolicySet{n}} + \vSetCardinality{\vQuery{}{j}})
\end{equation}

\subsection{Computing Optimal Regeneration Rate}

\oursystem will be able to cut the total cost for a querier which includes query evaluation and guard generation following the optimal regeneration rate. The cost of generating the guarded expression is proportional to the number of policies for the querier ($\vPolicySet{n}$). Assuming $k$ policies belonging to the querier are newly added since the guard ($\vGuard{}{}$) was last generated, we denote cost of guard generation by $\vGuardGen(\vPolicySet{n} + \vPolicySet{k})$. Given $\vDatabase$ and $\vUser{}{j}$ with $N$ policy insertions and $Q$ queries posed by $\vUser{}{j}$, the optimal number of policy insertions ($\vOptimalK$) before regenerating the guarded expression for $\vUser{}{j}$ is given by
\begin{equation}{\label{eq:guardMaintenance}}
\vOptimalK = \operatorname*{argmin}_{k \leq N} \sum_{i=1}^{\dfrac{N}{k}}( \vCostMethod{\vGuard{}{}, \vQuery{}{f(k)}, \vPolicySet{k}} + \vGuardGen(\vPolicySet{n} + \vPolicySet{k}))
\end{equation}

We divide that the total number of policies ($N$) into equal intervals of size $k$. To simplify the derivation, we assume that queries are uniform and the number of queries posed by the querier during that interval is given by $f(k)$. We define $f(k)$ based on $\vRateP$  which is the rate at which new policies are added (number of policies per unit time) and $\vRateQ$ which is the rate at which queries are posed by $\vUser{}{j}$ to $\vDatabase$. We combine both to define $\vRatePQ$ as the number of queries posed per policy insertion ($\dfrac{\vRateQ}{\vRateP}$) \footnote{We assume that $\vDatabase$ remains static which is only true for OLAP queries. Monitoring data insertion rate for each user will incur a significant overhead that will invalidate the usefulness of this approach.}. The number of queries during each interval of $\dfrac{N}{k}$ is given by $f(k) =  (j \mid 1 \leq j \leq k*\vRatePQ$. Finally, we simplify the guard generation cost as a constant ($\vGuardGen$) as it is dominated by the much larger $\vPolicySet{n}$. Putting all these together we have:
\begin{equation}{\label{eq:guardMaintenance}}
\vOptimalK = \operatorname*{argmin}_{k \leq N} \sum_{i=1}^{\dfrac{N}{k}} \biggl(\sum_{j=1}^{k*r_{qp}} \vCostMethod{\vGuard{}{}, \vQuery{}{j}, \vPolicySet{k}} + \vGuardGen \biggr)
\end{equation}

Expanding the first cost term with the cost of query evaluation from Equation~\ref{eq:guardQueryFinal} for insertion of k policies with the assumption that all queries are uniform and \\ $\vSelectivityMethod{\vObjectCondition{}{\vPolicySet{1}} \cup \vObjectCondition{}{\vPolicySet{2}} \cdots \cup \vObjectCondition{}{\vPolicySet{k}}} \subseteq \vSelectivityMethod{\vObjectCondition{}{\vGuard{}{}}}$
\begin{align}{\label{eq:forKpolicies}}
\mathclap{\vCostMethod{\vGuard{}{}, \vQuery{}{j}, \vPolicySet{k}}} & \notag \\ 
&= \vRatePQ.\vSelectivityMethod{\vObjectCondition{}{\vGuard{}{}}}.(\vReadCost + \vShortCircuit.\vEvalCost.(\vSetCardinality{\vPolicySet{n}} + \vSetCardinality{\vQuery{}{}})) \notag \\
&+ \vRatePQ.\vSelectivityMethod{\vObjectCondition{}{\vGuard{}{}}}.(\vReadCost + \vShortCircuit.\vEvalCost.(\vSetCardinality{\vPolicySet{n}} + 1 + \vSetCardinality{\vQuery{}{}})) \notag \\
&+ \cdots \notag \\
&+  \vRatePQ.\vSelectivityMethod{\vObjectCondition{}{\vGuard{}{}}}.(\vReadCost + \vShortCircuit.\vEvalCost.(\vSetCardinality{\vPolicySet{n}} + k + \vSetCardinality{\vQuery{}{}})) \notag \\
&= k.\vRatePQ.\vSelectivityMethod{\vObjectCondition{}{\vGuard{}{}}}.\vReadCost \notag \\
&+ \vRatePQ.\vSelectivityMethod{\vObjectCondition{}{\vGuard{}{}}}.\vEvalCost.\vShortCircuit.(k.\vSetCardinality{\vQuery{}{}} + k.\vSetCardinality{\vPolicySet{n}} + \dfrac{k.(k-1)}{2}) 
\end{align}

Using this equation in the previous minimization and replacing the summations with uniformity assumptions, the $\vOptimalK$ is given by 
\begin{align}
\vOptimalK = \operatorname*{argmin}_{k \leq N} \dfrac{N}{k}.\biggl(k.\vRatePQ.\vSelectivityMethod{\vObjectCondition{}{\vGuard{}{}}}. \\(\vReadCost + \vEvalCost.\vShortCircuit.(\vSetCardinality{\vQuery{}{}} + \vSetCardinality{\vPolicySet{n}} + \dfrac{(k-1)}{2}))\biggr) \notag
\end{align}

As our goal is to find the minimal k, we take the derivative of the above with respect to k and set it equal to 0.
\begin{align}
    \dfrac{\vSelectivityMethod{\vObjectCondition{}{\vGuard{}{}}}.\vShortCircuit.\vEvalCost.\vRatePQ}{2} - \dfrac{2.\vGuardGen}{k^2} = 0 \notag
\end{align}

Simplifying it for k, we have 
\begin{equation}{\label{eq:optimalK}}
    k = \sqrt{\dfrac{4.\vGuardGen}{\vSelectivityMethod{\vObjectCondition{}{\vGuard{}{}}}.\vShortCircuit.\vEvalCost.\vRatePQ}}
\end{equation}

The second derivative with respect to k is a positive value and therefore the k value derived by Equation~\ref{eq:optimalK} minimizes the cost of query evaluation and guard generation for $\vUser{}{j}$. Based on the simplifying assumptions used in this derivation, $\vOptimalK$ is an upper bound on the number of policy insertions before the guarded expression is updated. We now prove when it is most beneficial to regenerate the guarded expression after the insertion of $k^th$ policy.

\begin{theorem}
If the optimal rate of guard regeneration is set to k policies as in Equation~\ref{eq:optimalK}, then it is best to regenerate immediately after the $k^{th}$ policy has arrived.
\end{theorem}

We prove this by contradiction. Assuming the guard regeneration rate is set to k policies for a querier and we regenerate $\vGuard{}{}$ at k + $\delta$. If $\delta > \dfrac{1}{r_p}$, then regeneration rate is set at k + 1 and not k which is a contradiction. If $\delta > \dfrac{1}{r_q}$, then the new query will be evaluated using $\vGuard{}{}$ and the set of k policies which is higher compared to using the regenerated guarded expression as shown in the derivation above. Therefore $\delta < \dfrac{1}{r_p}$ and $\delta < \dfrac{1}{r_q}$ and thus regenerating immediately after $k^{th}$ policy will minimize the cost. \qed

\fi

\section{Experimental Evaluation}
\label{sect:exp}

\subsection{Experimental Setup}
\label{sect:experimentalSetup}

\vspace{0.1cm}
\noindent
\textbf{Datasets}. We used the {\em TIPPERS} dataset~\cite{mehrotra2016tippers} consisting of connectivity logs generated by the 64 WiFi Access Points (APs) at the Computer Science building at UC Irvine for a period of three months. These logs are generated when a WiFi enabled device (e.g., a smartphone or tablet) connects to one of the WiFi APs and contain the hashed identification of the device's MAC, the AP's MAC, and a timestamp. The dataset comprises~3.9M~events corresponding to~36K~different devices (the signal of some of the WiFi APs bleeds to outside the building and passerby devices/people are also observed). This information can be used to derive the occupancy levels in different parts of the building and to provide diverse location-based services (see Section~\ref{sect:caseStudy}) based on since device MACs can be used to identify individuals. Since location information is privacy-sensitive, it is essential to limit access to this data based on individuals' preferences. 
\ifextended
Table~\ref{tippers-schema} shows the schema of the different tables in the {\em TIPPERS} dataset. \textit{WiFi\_Dataset} stores the logs generated at each \textit{WiFi\_AP} when the devices of a \textit{User} connects to them. \textit{User\_Group} and \textit{User\_Group\_Membership} keeps track of the groups and their members respectively.

\begin{table}[!htb]
\scriptsize
\centering
\caption{TIPPERS data schema.} \label{tippers-schema}
\begin{tabular}{l l l}  
      Table           & Columns  &  Data type \\ \hline
\multirow{3}{*}{Users} & id & int  \\  \cline{2-3} 
                  & device & varchar \\  \cline{2-3} 
                  & office &  int \\ \hline
\multirow{3}{*}{User\_Groups} & id  & int \\ \cline{2-3} 
                  & name & varchar  \\ \cline{2-3} 
                  & owner & varchar  \\ \hline
\multirow{2}{*}{User\_Group\_Membership} & user\_group\_id &  int \\ \cline{2-3} 
                  & user\_id & int  \\ \hline
\multirow{3}{*}{Location} & id  & int  \\ \cline{2-3} 
                  & name & varchar  \\ \cline{2-3} 
                  & type & varchar \\  \hline
\multirow{5}{*}{WiFi\_Dataset} & id  & int  \\ \cline{2-3} 
                  & wifiAP & int  \\ \cline{2-3} 
                  & owner  & int \\ \cline{2-3} 
                  & ts-time & time \\ \cline{2-3} 
                  & ts-date & date
\end{tabular}
\end{table}

\fi

We also used a synthetic dataset containing WiFi connectivity events in a shopping mall for scalability experiments with even larger number of policies. We refer to this dataset as {\em Mall}. We generated the {\em Mall} dataset using the IoT data generation tool in~\cite{iotbenchmark} to generate synthetic trajectories of people in a space (we used the floorplan of a mall extracted from the Web) and sensor data based on those. The dataset contains 1.7M events from 2,651 different devices representing customers.
\ifextended
Table~\ref{mall-schema} shows the schema of the tables in the {\em Mall} dataset.  

\begin{table}[!htb]
\scriptsize
\centering
\caption{Mall data schema.} \label{mall-schema}
\begin{tabular}{l l l}  
      Table           & Columns  &  Data type \\ \hline
\multirow{3}{*}{Users} & id & int  \\  \cline{2-3} 
                  & device & varchar \\  \cline{2-3} 
                  & interest &  varchar \\ \hline
\multirow{3}{*}{Shop} & id  & int  \\ \cline{2-3} 
                  & name & varchar  \\ \cline{2-3} 
                  & type & varchar \\  \hline
\multirow{5}{*}{WiFi\_Connectivity} & id  & int  \\ \cline{2-3} 
                  & shop\_id & int  \\ \cline{2-3} 
                  & owner  & int \\ \cline{2-3} 
                  & obs\_time & time \\ \cline{2-3} 
                  & obs\_date & date
\end{tabular}
\end{table}

\fi

\vspace{0.1cm}
\noindent
\textbf{Queries}. We used a set of query templates based on the recent IoT SmartBench benchmark~\cite{iotbenchmark} which include a mix of analytical and real-time tasks and target queries about (group of) individuals. 
Specifically, query templates $Q_1$ - Retrieve the devices connected for a list of locations during a time period (e.g., for location surveillance); $Q_2$ - Retrieve devices connected for a list of given MAC addresses during a time period (e.g., for device surveillance); $Q_3$ - Number of devices from a group or profile of users in a given location (e.g., for analytic purposes). 
\ifextended
The SQL version of the queries is thus:
\begin{lstlisting}
Q1=(SELECT * FROM WiFi_Dataset AS W 
    WHERE W.wifiAP IN ($[ap]$) W.ts-time BETWEEN $t1$ AND $t2$ AND W.ts-date BETWEEN $d1$ AND $d2$)
Q2=(SELECT * FROM WiFi_Dataset AS W 
    WHERE W.owner in ($[devices]$) AND W.ts-time BETWEEN $t1$ AND $t2$ AND W.ts-date BETWEEN $d1$ AND $d2$)
Q3=(SELECT * FROM WiFi_Dataset AS W, User_Group_Membership AS UG 
    WHERE UG.user_group_id = $group-id$ 
    AND UG.user_id = W.owner AND W.ts-time BETWEEN $t1$ AND $t2$ AND W.ts-date BETWEEN $d1$ AND $d2$)
\end{lstlisting}
\fi

Based on these templates, we generated queries at three different selectivities (low, medium, high) by modifying configuration parameters (locations, users, time periods).
Below, when we refer to a particular query type (i.e., $Q_1$, $Q_2$, or $Q_3$) we will refer to the set of queries generated for such type.

\vspace{0.1cm}
\noindent
\textbf{Policy Generation}.
The TIPPERS dataset, collected for a limited duration with special permission from the University for the purpose of research, does not include user-defined policies. We therefore generated a set of synthetic policies. As part of the TIPPERS project, we conducted several town hall meetings and online surveys to understand the privacy preferences of users about sharing their WiFi-based location data. 
The surveys, as well as prior research~\cite{DBLP:conf/percom/0001K17,lin2014privacy}, indicate that users express their privacy preferences based on different user profiles (e.g., students for faculty) or groups (e.g., my coworkers, classmates, friends, etc.). Thus, we used a profile-based approach to generate policies specifying which events belonging to  individual can be accessed by a given querier (based on their profile) for a specific purpose in a given context (e.g., location, time).

We classified devices in the TIPPERS dataset as belonging to users with different profiles (denoted by $\vProfileMethod{\vUser{}{k}}$ for User~$\vUser{}{k}$) based on the total time spent in the building and connectivity patterns. Devices which rarely connect to APs in the building (i.e., less than~5\% of the days) are classified as \textit{visitors}. The non-visitor devices are then classified based on the type of rooms they spent most time in: \textit{staff} (staff offices), \textit{undergraduate students} (classrooms), \textit{graduate students} (labs), and \textit{faculty} (professor offices). As a result, we classified 31,796 visitors, 1,029 staff, 388 faculty, 1,795 undergraduate, and 1,428 graduate from a total of 36,436 unique devices in the dataset. Our classification is consistent with the expected numbers for the population of the monitored building. We also grouped users into groups based on the affinity of their devices to rooms in the building which is defined in terms of time spent in each region per day. Thus, each device is assigned to a group with maximum affinity. In total, we generated 56 groups with an average of 108 devices per group. 

We define two kinds of policies based on whether they are an unconcerned user or an advanced user as described in Section~\ref{sect:caseStudy}. Unconcerned users subscribe to the default policies set by administrator which allows access to their data based on user-groups and profiles. Given the schema in Table~\ref{tippers-schema} and the unconcerned user $\vUser{}{k}$ we generate the following default policies:

\ifextended
\squishlist
    \item Data associated with $\vUser{}{k}$ collected during working hours can be accessed by members of $\vGroupMethod{\vUser{}{k}}$.
    \item Data associated with $\vUser{}{k}$ collected at any time can be accessed by overlapping members of $\vGroupMethod{\vUser{}{k}}$ and $\vProfileMethod{\vUser{}{k}}$. 
\squishend

On the other hand, an advanced user define on average 40 policies, given the large number of control options (such as device, time, groups, profiles, and locations) in our setting. In total, the policy dataset generated contains 869,470 policies with each individual defining 472 policies on average and appearing as querier in 188 policies defined by others on average. The above policies are defined to allow access to data in different situations. Any other access that is not captured by the previous policies will be denied (based on the default opt-out semantics defined in Section~\ref{sect:modeling}). 

\ifextended

\else

\fi

\ifextended

Table~\ref{policy-table} shows the schema and several sample policies generated for three different queriers. The \textit{inserted\_at} and \textit{action} columns are skipped for brevity. Table~\ref{oc-table} shows the corresponding object conditions which are part of two policies.

\begin{table}[!htb]
\small
\centering
\caption{Policy Table} \label{policy-table}
 \begin{tabular}{c l l l} 
 id & table & querier & purpose \\ [0.5ex] 
 \hline
 1 & WiFi\_Dataset & Prof.John Smith & Attendance   \\ 
 \hline
 2 & WiFi\_Dataset & Bob Belcher & Lunch Group  \\ 
 \hline
 3 & WiFi\_Dataset & Prof.John Smith & Attendance  \\ 
 \hline
 4 & WiFi\_Dataset & Liz Lemon & Project Group  \\ 
 \hline
 5 & WiFi\_Dataset & Prof.John Smith & Attendance 
\end{tabular}
\end{table}

\begin{table}[!htb]
\small
\centering
\caption{Policy Object Conditions Table} \label{oc-table}
 \begin{tabular}{c c l l c r} 
 id & policy\_id & attr\_type & attr & op & val \\ [0.5ex] 
 \hline
 1 & 1 & int & owner & $=$ & 120   \\ 
 \hline
 2 & 1 & time & ts-time & $\geq$ & 09:00:00     \\ 
 \hline
 3 & 1 & time & ts-time & $\leq$ & 10:00:00     \\ 
 \hline
 4 & 1 & int & wifiAP & $=$ & 1200     \\ 
 \hline
 5 & 2 & int & owner &  $=$ & 145  \\ 
 \hline
 6 & 2 & int & wifiAP & $=$ & 2300
\end{tabular}
\end{table}

\fi

For the \textit{Mall} dataset, the shops were categorized into six types based on the services they provide, e.g., arcade, movies, etc.  We also classified customers into regular and irregular, based on their shop visits. For each customer, we then defined two types of policies depending on whether they were regular or irregular. Regular customers allowed shops they visit the most to have access to their location during open hours. On the other hand, irregular customers shared their data only with specific shop types depending on if there were sales or discounts. Finally, if a customer expressed an interest in a particular shop category, we also generated policies which allowed access of their data to the shops in the category for a short period of time (e.g., lightning sales). In total, this policy dataset generated on top of {\em Mall} dataset contains in 19,364 policies defined for 35 shops (queriers) in the mall with 551 policies on average per shop.

\vspace{0.1cm}
\noindent
\textbf{Database System}. We ran the experiments on an individual machine (CentOS 7.6, Intel(R) Xeon(R) CPU E5-4640, 2799.902 Mhz, 20480 KB cache size) in a cluster with a shared total memory of 132 GB. 
We performed experiments on MySQL 8.0.3 with InnoDB as it is an open source DBMS which supports index usage hints. We configured the buffer\_pool\_size to 4 GB. We also performed experiments on PostgreSQL 13.0 with shared\_buffers configured to 4 GB.

\subsection{Experimental Results}

We first study the speed and quality of guarded expression generation algorithm for the policy dataset generated based on \textit{TIPPERS} (Experiment 1). Secondly, we performed experiments to validate the design choices in \oursystem (Experiment 2) as well as compare the performance of \oursystem against the baselines (Experiment 3). Fourth, we tested \oursystem on PostgreSQL which does not support index usage hints and is therefore different from MySQL on which all the previous experiments were run (Experiment 4). In the final experiment, we stress test our approach with a very large number of policies (Experiment 5). The first four experiments were done on \textit{TIPPERS} and the final experiment was done on \textit{Mall} dataset.

\setlength{\intextsep}{2pt}%
\setlength{\columnsep}{12pt}%
\begin{wrapfigure}{R}{0.25\textwidth}
\centering
\hspace{6mm}
	\includegraphics[width=0.25\textwidth]{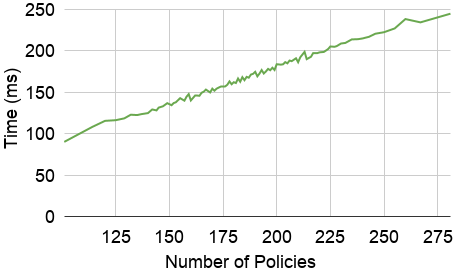}
\caption{Guard generation cost.}
\label{fig:guardGenerationTest}
\end{wrapfigure}

\vspace{0.1cm}
\noindent
\textbf{Experiment 1: Cost for generating Guarded Expressions and Effectiveness.} The goal of this experiment is to study the cost of generating guarded expressions for a querier as factor of the number of policies and the quality of generated guards. For analyzing the cost of guarded expression generation, we generate guarded expressions for all the users using the algorithm described in Section~\ref{sect:guardSel} and collect the generation times in a set. We sort these costs (in milliseconds) of generating guarded expressions for different users and average the value for every group of 50 users showing the result in Figure~\ref{fig:guardGenerationTest}. 
The cost of guard generation increases linearly with number of policies. 
As guarded expression generation is also dependent on the selectivity of policies, number of candidate guards generated, which is also a factor of overlap between predicates, we sometimes observe a slight decrease in the time taken with increasing policies. The overhead of the cost of generating guarded expression is minimal, for instance, the cost of generating a guard for a querier with 160 policies associated (e.g., the student trying to locate classmates explained in Section~\ref{sect:caseStudy}) is around 150ms.

\begin{center}
\begin{minipage}{\linewidth}
    \begin{minipage}[b]{0.54\linewidth}
        \vspace{0pt}
        \centering
        \scriptsize
        \begin{tabular}{lllll}
            & min  & avg  & max  & SD  \\ \hline
            $\vSetCardinality{\vPolicy{}{\vUser{}{k}}}$ & 31 & 187 & 359 & 38 \\\hline
            $\vSetCardinality{\vGuard{}{}}$ & 2 & 31 & 60 & 10  \\\hline
            $\vSetCardinality{\vPolicy{}{\vGuard{}{i}}}$ & 4  & 7 & 60 & 5  \\\hline
            $\vSelectivityMethod{\vGuard{}{i}}$ & 0.01\%  & 3\% & 24\%  & 2\%  \\\hline
            Savings  & 0.99  & 0.99 & 1  & $7\mathrm{e}{-4}$ \\ \\
        \end{tabular}
        \captionof{table}{Analysis of policies and generated guards.}
        \label{tab:guardAnalysis}
    \end{minipage}
    \hfill
    \begin{minipage}[b]{0.4\linewidth}
        \vspace{0pt}
        \centering
        \scriptsize
        \begin{tabular}{lrr}
            \multirow{2}{*}{$\vSelectivityMethod{\vGuard{}{}}$} & \multicolumn{2}{c}{$\vSetCardinality{\vGuard{}{}}$} \\
            \cmidrule(lr){2-3} & low   & high    \\ \hline
            low                & 227.2 & 537.0   \\ \hline
            high               & 469.0 & 1,406.7 \\ \hline \\ \\
        \end{tabular}
        \captionof{table}{Analysis of number of guards and total cardinality.}
        \label{tab:guardsNumbervsCardinality}
    \end{minipage}
\end{minipage}
\end{center}

We present the results of analyzing the policies and guarded expressions in Table~\ref{tab:guardAnalysis}. Each user on average have defined 187 policies ($\vSetCardinality{\vPolicy{}{\vUser{}{k}}}$). This number of policies depends on their profiles (e.g., student) and group memberships. \oursystem creates an average of 31 guards per user with the mean partition cardinality (i.e., $\vSetCardinality{\vPolicy{}{\vGuard{}{i}}}$) as 7. The total cardinality of  guards in the guarded expression is low (i.e., $\vSelectivityMethod{\vGuard{}{i}}$) which helps in filtering out tuples before performing policy evaluation. In cases with high cardinality guards (e.g., maximum of 24\%), \oursystem will not use force an index scan in that particular guard as explained in Section~\ref{sect:implementingSieve}.  Savings is computed as ratio of the difference between total number of policy evaluations without and with using the guard and the number of policy evaluations. This was computed on a smaller sample of the entire dataset and the results show guards help in eliminating around 99\% of the policy checks compared to policy evaluation.

\vspace{0.1cm}
\noindent
\textbf{Experiment 2.1: Inline vs. Operator $\Delta$.}
SIEVE uses a cost model to determine for each guard whether to inline the policies or to evaluate the policies using the $\Delta$ operator. The $\Delta$ operator has an associated overhead of UDF invocation but it can utilize the tuple context to reduce the number of policies that needs to be checked per tuple. For the purpose of studying this tradeoff in both inlining and using the $\Delta$ operator, we gradually increased the number of policies that are part of the partition of a guard and observed the cost of policy evaluation. As expected, we observed that when the number of policies are about 120, the cost of UDF invocations is amortized by the savings from filtering policies by the $\Delta$ operator (see Figure~\ref{fig:designChoice1}). 

\begin{center}
\begin{minipage}{\linewidth}
    \begin{minipage}[b]{0.49\linewidth}
        \vspace{0pt}
        \centering
        \includegraphics[width=\columnwidth]{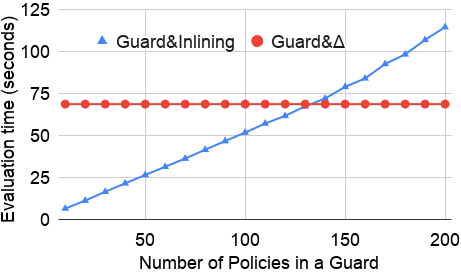}
        \captionof{figure}{$Inlining$ vs. $\Delta$.}
        \label{fig:designChoice1}
    \end{minipage}
    \hfill
    \begin{minipage}[b]{0.5\linewidth}
        \vspace{0pt}
        \centering
        \includegraphics[width=\columnwidth]{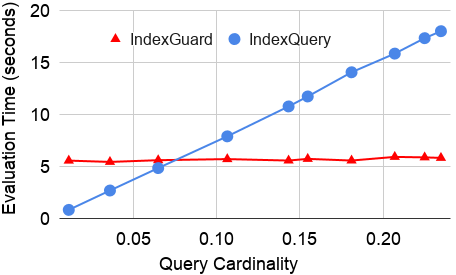}
        \captionof{figure}{Index choice.}
        \label{fig:designChoice2}
  \end{minipage}
\end{minipage}
\end{center}

\vspace{0.1cm}
\noindent
\textbf{Experiment 2.2: Query Index vs. Guard Index.}
In \oursystem, we use a cost model to choose between using the  \textit{IndexQuery} and \textit{IndexGuards} as explained in Section~\ref{sect:implementingSieve}. We evaluated this cost model by analyzing the cost of evaluation against increasing query cardinality for three different guard cardinalities (low, medium, high). Figure~\ref{fig:designChoice2} shows the results averaged across these three guard cardinalities. As expected, at low query cardinality it is better to utilize \textit{IndexQuery}, while at medium and high query cardinalities ($>0.07$), \textit{IndexGuards} are the better choice. Note that in both these options, guarded expressions are used as filter on top of the results from Index Scan.

\vspace{0.1cm}
\noindent 
\textbf{Experiment 3: Query Evaluation Performance.} We compare the performance of \oursystem (implemented as detailed in Section~\ref{sect:implementingSieve}) against three different baselines. In the first baseline, {\vBaselineP}, we append the policies that apply to the querier to the |WHERE| condition of the query.
Second, {\vBaselineI}, performs an index scan per policy (forced using index usage hints) and combines the results using the |UNION|. Third, {\vBaselineU} is similar to {\vBaselineP} but instead of using the policy expression, it uses a UDF defined on the relation to evaluate the policies. The UDF takes as input all the attributes of the tuple. {\vBaselineU} reduces significantly the number of policies to be evaluated per tuple and is therefore an interesting optimization strategy for low cardinality queries. UDF invocations are expensive, therefore so it might be preferable to execute the UDF as late as possible from the optimization perspective~\cite{hellerstein1998optimization}. To preserve correctness of policy enforcement as defined in Section~\ref{sect:modeling}, UDF operations have to be performed before any non-monotonic query operations.

For each of the query types ($Q1$, $Q2$, $Q_3$), we generate a workload of queries with three different selectivity classes posed by five different queriers of belonging to four different profiles. The values chosen for these three selectivity classes (low, medium, high) differed depending upon the query type. We execute each query along with the access control mechanism~5~times and average the execution times. The experimental results below give the average warm performance per query. The time out was set at 30 seconds. If a strategy timed out for all queries of that group we show the value $TO$. 
If a strategy timed out for some of the queries in a group but not all, the table shows the average performance only for those queries that were executed to completion; those time values are denoted as $t^{+}$.

\begin{table}[!htb]
\caption{Overall comparison of performance for $Q1$, $Q2$, and $Q3$ (in ms).}
\label{tab:performanceAllQueries}
\centering
\scriptsize
    \begin{tabular}{lcrrrr}
         &  $\vSelectivityMethod{\vQuery{}{}}$ & \vBaselineP & \vBaselineI & \vBaselineU & \oursystem \\ \hline
        \multirow{3}{*}{$Q1$} & low  & 1,668        & 906   & 9,122        & 418 \\ 
                              & mid  & 15,356       & 910   & 23,575$^{+}$ & 453 \\ 
                              & high & TO           & 937   & TO           & 523 \\ \hline
        \multirow{3}{*}{$Q2$} & low  & 860          & 916   & 7,787        & 407 \\ 
                              & mid  & 7,191        & 922   & 22,617$^{+}$ & 454 \\ 
                              & high & 29,765$^{+}$ & 962   & TO           & 475 \\ \hline  
        \multirow{3}{*}{$Q3$} & low  & 883          & 881   & 14,379        & 477 \\ 
                              & mid  & 2,217          & 2,209   & TO & 476 \\ 
                              & high & 3,502        & 3,543 & TO           & 521 \\
    \end{tabular}
\end{table}

Table~\ref{tab:performanceAllQueries} shows the average performance for the three query types. Performance of {\vBaselineP} and {\vBaselineU} degrades with increasing cardinality of the associated query as they rely on the query predicate for reading the tuples. The relative reduction in overhead for Q3 for {\vBaselineP} at high cardinalities is because the optimizer is able to use the low cardinality join condition to perform a nested index loop join. \oursystem and \vBaselineI performance stays the same across query cardinalities as they utilize the policy and guard predicates for reading the tuples and is therefore not affected by the query cardinality. The increase in the speedup between these two sets of approaches clearly demonstrate that exploiting indices paid off. For {\vBaselineP}, the optimizer is not able to exploit indices at high cardinalities and resorts to performing linear scan. In {\vBaselineU}, the cost of UDF invocation per tuple far outweighed any benefits from filtering of policies. {\vBaselineI} generated by careful rewriting with an index scan per policy, performs significantly better than the previous two baselines. The performance degrade of {\vBaselineI} for Q3 is due to the optimizer preferring to perform the nested loop join first instead of the index scans. In comparison to all these baselines, {\oursystem} is significantly faster at all different query cardinalities. 

\ifextended
The extended results for $Q1$, $Q2$, and $Q3$ by querier profile (F-Faculty, G-Graduate, U-Undergrad, and S-Staff are shown in Table~\ref{tab:performanceQ1}, Table~\ref{tab:performanceQ2}, and Table~\ref{tab:performanceQ3}, respectively.
\fi

\ifextended

\begin{table}[!htb]
\caption{Comparison of performance for $Q1$ (in ms).}
\label{tab:performanceQ1}
\centering
\scriptsize
    \begin{tabular}{lcrrrr}
        Pr. &  $\vSelectivityMethod{\vQuery{}{}}$ & \vBaselineP & \vBaselineI & \vBaselineU & \oursystem \\ \hline
        \multirow{3}{*}{F} & l & 1,560  & 972   & 9,398        & 357 \\ 
                           & m & 14,533 & 949   & 23,362$^{+}$ & 352 \\ 
                           & h & TO     & 962   & TO           & 413 \\ \hline
        \multirow{3}{*}{G} & l & 1,794  & 998   & 9,573        & 426 \\ 
                           & m & 16,737 & 994   & 23,735$^{+}$ & 495 \\ 
                           & h & TO     & 990   & TO           & 565 \\ \hline
        \multirow{3}{*}{U} & l & 1,618  & 681   & 9,661        & 362 \\ 
                           & m & 15,432 & 751   & 23,692$^{+}$ & 394 \\ 
                           & h & TO     & 720   & TO           & 422 \\ \hline
        \multirow{3}{*}{S} & l & 1,701  & 975   & 7,854        & 526 \\ 
                           & m & 14,722 & 946   & 23,511$^{+}$ & 571 \\ 
                           & h & TO     & 1,077 & TO           & 691 \\ \hline
    \end{tabular}
\end{table}

\begin{table}[!htb]
\caption{Comparison of performance (in ms) for $Q2$.}
\label{tab:performanceQ2}
\centering
\scriptsize
    \begin{tabular}{lcrrrr}
        Pr. &  $\vSelectivityMethod{\vQuery{}{}}$& \vBaselineP & \vBaselineI & \vBaselineU & \oursystem\\ \hline
        \multirow{3}{*}{F} & l & 822    & 961   & 7,655        & 354 \\ 
                           & m & 6,929  & 975   & 22,502$^{+}$ & 354 \\ 
                           & h & 26,397 & 991   & TO           & 362 \\ \hline
        \multirow{3}{*}{G} & l & 947    & 1,009 & 8,084        & 404 \\ 
                           & m & 7,806  & 1,028 & 22,676$^{+}$ & 506 \\ 
                           & h & TO     & 1,080 & TO          & 537 \\ \hline
        \multirow{3}{*}{U} & l & 848    & 739   & 8,336        & 380 \\ 
                           & m & 7,156  & 725   & 22,863$^{+}$ & 368 \\ 
                           & h & TO     & 769   & TO           & 399 \\ \hline
        \multirow{3}{*}{S} & l & 822    & 954   & 7,073        & 489 \\ 
                           & m & 6,874  & 960   & 22,425$^{+}$ & 589 \\ 
                           & h & 28,347 & 1,007 & TO           & 603 \\ \hline
    \end{tabular}
\end{table}

\begin{table}[!htb]
\caption{Comparison of performance (in ms) for $Q3$.}
\label{tab:performanceQ3}
\centering
\scriptsize
    \begin{tabular}{lcrrrr}
        Pr. &  $\vSelectivityMethod{\vQuery{}{}}$& \vBaselineP & \vBaselineI & \vBaselineU & \oursystem\\ \hline
        \multirow{3}{*}{F} & l & 892   & 871   & 14,279        & 372 \\ 
                           & m & 2,302   & 2,248   & TO & 379 \\ 
                           & h & 3,595 & 3,662 & TO           & 405 \\ \hline
        \multirow{3}{*}{G} & l & 893   & 886   & 13,287         & 524 \\ 
                           & m & 2,183   & 2,200   & TO & 518 \\ 
                           & h & 3,486 & 3,487 & TO           & 568 \\ \hline
        \multirow{3}{*}{U} & l & 881   & 884   & 10,601        & 619 \\ 
                           & m & 2,174   & 2,200   & TO & 613 \\ 
                           & h & 3,512 & 3,446 & TO           & 668 \\ \hline
        \multirow{3}{*}{S} & l & 865   & 885   & 11,947        & 319 \\ 
                           & m & 2,211   & 2,188   & TO & 392 \\ 
                           & h & 3,512 & 3,576 & TO           & 444 \\ \hline
    \end{tabular}
\end{table}

\fi

\begin{center}
\begin{minipage}{\linewidth}
    \begin{minipage}[b]{0.49\linewidth}
        \vspace{0pt}
        \centering
        \includegraphics[width=\columnwidth]{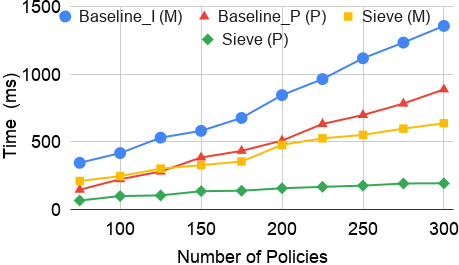}
        \captionof{figure}{\oursystem on MySQL and PostgreSQL.}
        \label{fig:exp4}
    \end{minipage}
    \hfill
    \begin{minipage}[b]{0.49\linewidth}
        \vspace{0pt}
        \centering
        \includegraphics[width=\columnwidth]{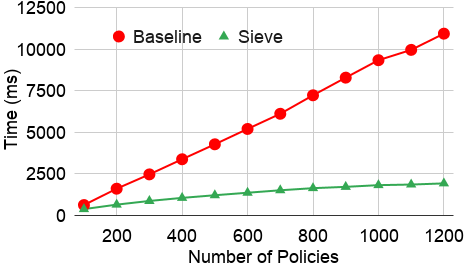}
        \captionof{figure}{Scalability comparison.}
        \label{fig:expMall}
    \end{minipage}
\end{minipage}
\end{center}

\vspace{0.1cm}
\noindent 
\textbf{Experiment 4: \oursystem on PostgreSQL}.  
In the previous experiments we used MySQL, which supports hints for index usage, thus enabling SIEVE to explicitly force the optimizer to choose guard indexes. However, other DBMSs, such as PostgreSQL, do not support index usage hints explicitly (as discussed in Section~\ref{sect:implementingGuards}). To study \oursystem's performance in such systems, we implemented a \oursystem connector to PostgreSQL using the same rewrite strategy but without index usage hints. In order to have a cumulative set of policies (i.e., the larger set of policies contain the smaller set of policies) for evaluation, we chose 5 queriers with at least 300 policies in the dataset. For each querier, we divided their policies into 10 different sets of increasing number of policies starting with smallest set of 75 policies. The order and the specific policies in these sets were varied 3 times by random sampling. The results in Figure~\ref{fig:exp4} shows the average performance of different strategies for each set size averaged across queriers and the samples for SELECT ALL queries.

The four strategies tested in this experiment are: best performing baseline for MySQL (\vBaselineI (M)) from Experiment 3, the baseline in PostgreSQL (\vBaselineP (P)), and \oursystem in both MySQL and PostgreSQL (\oursystem(M) and \oursystem(P)). The results show that not only \oursystem outperforms the baseline in PostgreSQL but also the speedup factor w.r.t. the baseline is even higher than in MySQL. Additionally, the speedup factor in PostgreSQL is the highest at largest number of policies. Based on our analysis of the query plan chosen by PostgreSQL, it correctly chooses the guards for performing index scans (as intended by \oursystem) even without the index usage hints. In addition, PostgreSQL supports combining multiple index scans by preparing a bitmap in memory\footnote{\url{https://www.postgresql.org/docs/12/indexes-bitmap-scans.html}}. It used these bitmaps to \textit{OR} the results from the guards whenever it was possible, and the only resultant table rows are visited and obtained from the disk. With a larger number of guards (for larger number of policies), PostgreSQL was also able to more efficiently filter out tuples compared to using the policies. Thus, {\oursystem} benefits from reduced number of disk reads (due to bitmap) as well as a smaller number of evaluations against the partition of the guarded expression.

\vspace{0.1cm}
\noindent 
\textbf{Experiment 5: Scalability.}
The previous experiment shows that the speedup of \oursystem w.r.t. the baselines increases with an increasing number of policies, especially for PostgreSQL. We explore this aspect further on PostgreSQL using the {\em Mall} dataset where the generation of very large number of policies is more feasible as we can simulate more customers. 
We used the same process than in Experiment 4 to generate cumulative set of policies by choosing 5 queriers/shops with at least 1,200 policies defined for them. Figure~\ref{fig:expMall} reaffirms how the speedup of \oursystem compared against the baseline increases linearly starting from a factor of 1.6 for 100 policies to a factor of 5.6 for 1,200 policies. We analyzed the query plan selected by the optimizer for the \oursystem rewritten queries. We observed that with larger number of guards, PostgreSQL is able to utilize the bitmaps in memory to gain additional speedups from guarded expressions (as explained in Experiment 4). Also, this experiment shows that \oursystem outperforms the baseline for a different dataset which shows the generality of our approach.

\section{Conclusion}
\label{sect:conclusion}

In this paper, we presented \oursystem, a layered approach to enforcing large number of fine-grained policies during query execution. 
\oursystem combines two optimizations: reducing the number of policies that need to be checked against each tuple, and reducing the number of tuples that need to be checked against complex policy expressions. %
\oursystem is designed as a general purpose middleware approach and we have layered it on two different DBMSs. The experimental evaluation, using a real dataset and a synthetic one, highlights that \oursystem enables existing DBMSs to perform efficient access control. \oursystem significantly outperforms existing strategies for implementing policies based on query rewrite. The speedup factor increases with increasing number of policies and \oursystem's query processing time remains low even for thousands of policies per query.

We plan to leverage the experience gained while developing \oursystem to pursue a tighter integration with the database query optimizer. Also, we plan to study possible mechanisms to combine certain amount of preprocessing at insertion time to simplify policy checking and guard evaluation at query time to extend \oursystem to co-optimize a query and policy workload.

\bibliographystyle{abbrv}
\bibliography{references}

\end{document}
\endinput